\documentclass[a4paper,11pt]{article}
\pdfoutput=1 % if your are submitting a pdflatex (i.e. if you have
\usepackage{jheppub} % for details on the use of the package, please
\usepackage[T1]{fontenc} % if needed
\RequirePackage[displaymath]{lineno} % Display line numbers

\usepackage{epsfig}
\usepackage{graphicx}% Include figure files
\usepackage{dcolumn}% Align table columns on decimal point
\usepackage{bm}% bold math
\usepackage{ltablex,booktabs}
\usepackage{overpic}
\usepackage{subfigure}
\usepackage{float}
\usepackage{color}
\usepackage{amsmath}
\usepackage{mathcomp}
\usepackage{mathrsfs}
\usepackage{multirow}
\usepackage{rotating}
\usepackage{amssymb}
\usepackage{gensymb}
\usepackage{amsmath}
\usepackage{tabularx}
\usepackage{epstopdf}
\title{Amplitude analysis and branching fraction measurement of the decay \boldmath $D_{s}^{+} \to K^+\pi^{+}\pi^{-}\pi^{0}$}
\collaborationImg{\includegraphics[width=0.15\textwidth, angle=90]{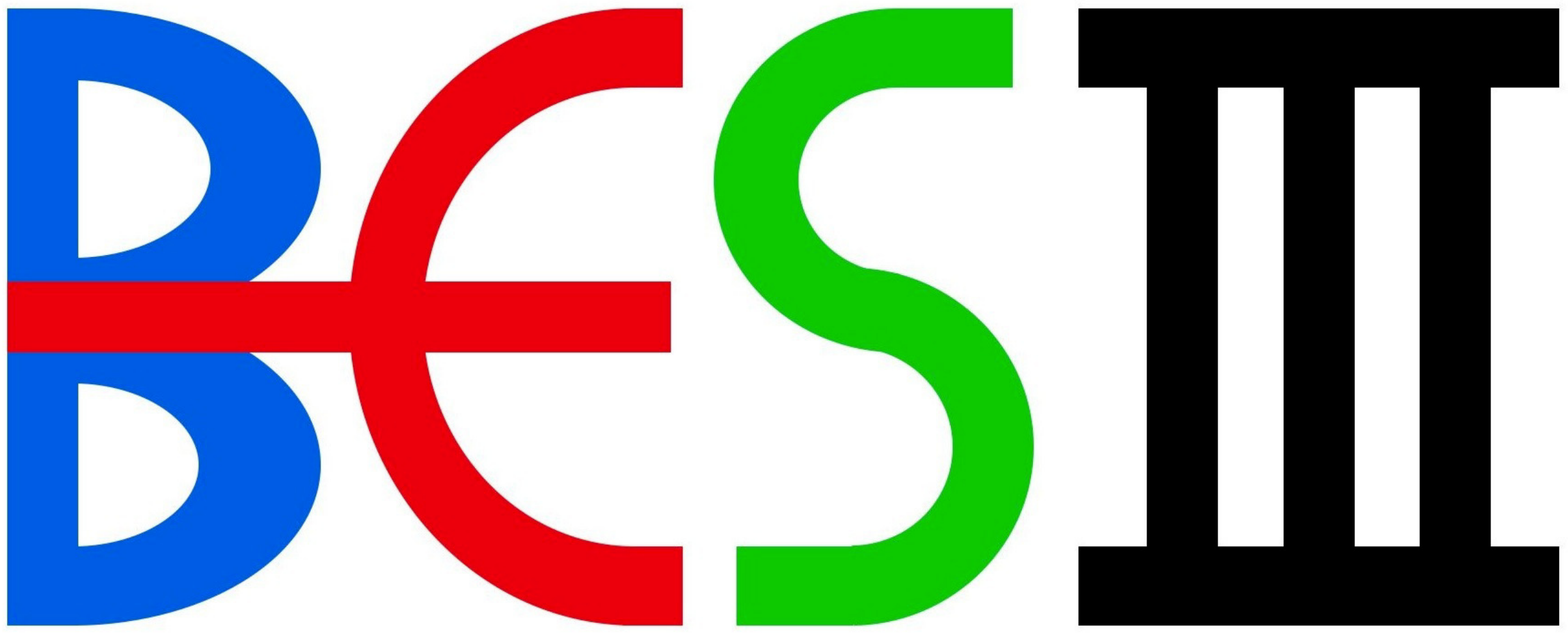}}
\collaboration{BESIII Collaboration}

\date{\today}
\abstract{
  The singly Cabibbo-suppressed decay $D_{s}^{+} \to K^+\pi^{+}\pi^{-}\pi^{0}$ is observed by using a data set corresponding to an integrated luminosity of
  6.32~$\rm fb^{-1}$ recorded by the BESIII detector at the centre-of-mass energies
  between 4.178 and 4.226~GeV. The first amplitude analysis of $D_{s}^{+} \to K^+\pi^{+}\pi^{-}\pi^{0}$
  reveals the sub-structures in this decay and determines the fractions and relative phases of different intermediate processes.
  The dominant intermediate process is $D_s^+ \to K^{*0}\rho^+$, with a fit fraction of
  $(40.5\pm2.8_{\rm{stat.}}\pm1.5_{\rm{syst.}})\%$.
  With the detection efficiency based on our amplitude analysis, the absolute branching fraction
  for $D_{s}^{+} \to K^+\pi^{+}\pi^{-}\pi^{0}$ is measured to be
  $(9.75\pm0.54_{\rm{stat.}}\pm0.17_{\rm{syst.}})\times 10^{-3}$.
}

\arxivnumber{}
\begin{document}
\maketitle
\flushbottom
%\normalsize
%\parskip=5pt plus 1pt minus 1pt

%------------------------------------------------------------------------------
\section{Introduction}
The hadronic decays of charmed mesons have been studied extensively in both experiment and theory since the discovery of charmed mesons in 1976 by Mark I~\cite{PRL37-255,PRL37-569}. However, a precise theoretical description for exclusive hadronic charmed meson decays is still challenging because the mass of charm quark is too light to adopt a sensible heavy quark expansion and too heavy to apply chiral perturbation theory~\cite{PRD81-074021}. Amplitude analyses and measurements of the branching fractions (BFs) for hadronic decays of charmed mesons provide valuable information about the underlying mechanism of the charmed meson decays.

\begin{figure*}[!htbp]
  \centering
  \includegraphics[width=0.495\textwidth]{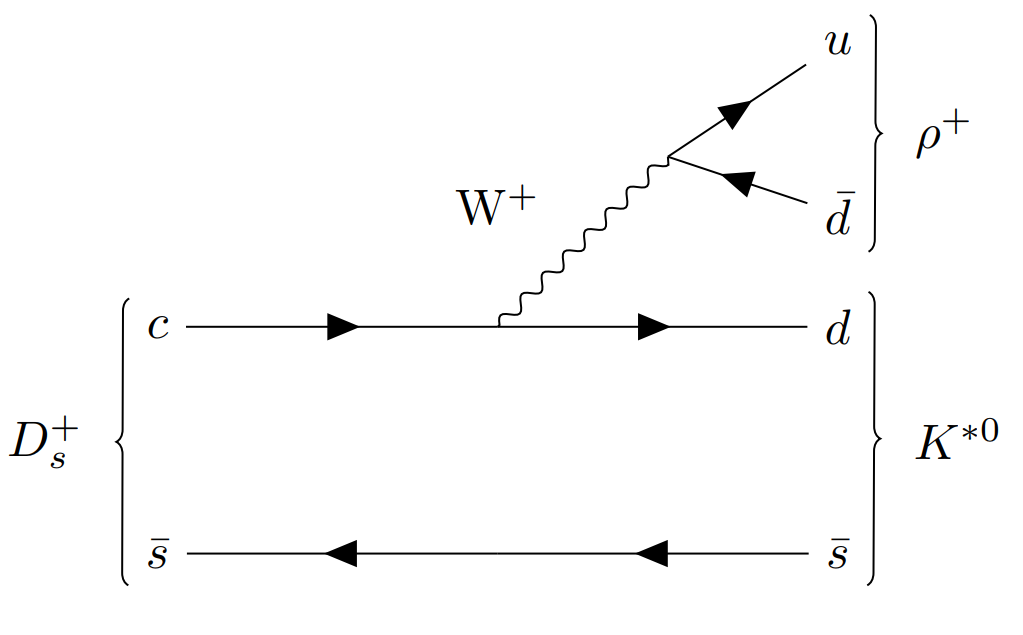}
  \includegraphics[width=0.495\textwidth]{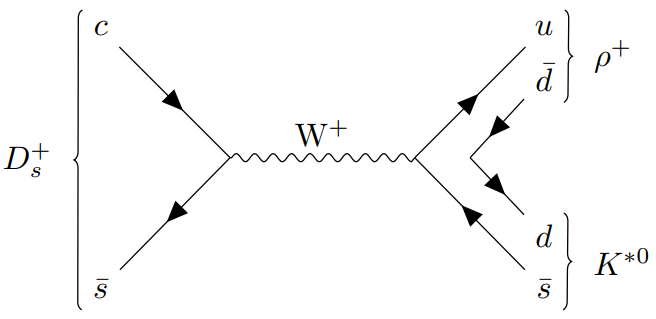}
    \caption{The $T$-diagrams~(left) and $A$-diagrams~(right) for the decay $D_s^+ \to K^{*0}\rho^+$
  } \label{fig:topology1}
\end{figure*}

\begin{figure*}[!htbp]
  \centering
  \includegraphics[width=0.495\textwidth]{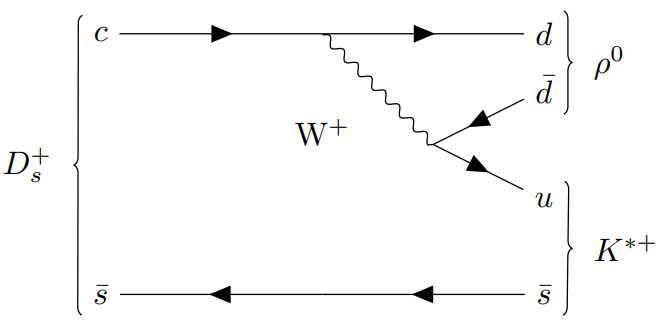}                                                 \includegraphics[width=0.495\textwidth]{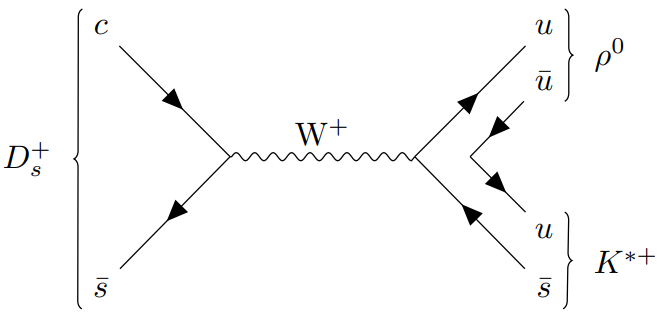}\\
  \caption{The $T$-diagrams~(left) and $A$-diagrams~(right) for the decay $D_s^+ \to K^{*+}\rho^0$.
  } \label{fig:topology2}
\end{figure*}

Four-body hadronic decays of $D^+_s$ mesons can be dominated by two-body intermediate processes~\cite{PDG}, such as $D^{+}_s \to VV$ and $D^{+}_s \to AP$ decays, where $V,\ A$, and $P$ denote vector, axial-vector and pseudoscalar mesons, respectively. The investigations of the $D^{+}_s \to VV$ decays have attracted a great deal of attention~\cite{VV1,VV2,VV3,VV4,PLB684-137}, but the experimental information about the $D^{+}_s \to VV$ decays is sparse. And the improved knowledge of BFs of $D^{+}_s \to AP$ decays, such as $D_{s}^{+} \to K_1(1270)^0\pi^{+}$ and $D_{s}^{+} \to K_1(1400)^0\pi^{+}$, is important to improve the understanding of the mixing of the $K_1(1270)^0$ and $K_1(1400)^0$ mesons~\cite{PLB707-116}.
The singly Cabibbo-suppressed hadronic decay of $D_{s}^{+} \to K^+\pi^{+}\pi^{-}\pi^{0}$ is expected to be dominated by the intermediate decays $D_s^+ \to K^{*0}\rho^+$ and $K_1^0\pi^+$ ($\rho$ denotes $\rho(770)$, $K^{*}$ denotes $K^{*}(892)$ and $K_1$ denotes $K_1(1270)/K_1(1400)$), since the decay width calculated by external W-emission process with final states of neutral kaonioc states (i.e. $K^{*0}$, $K_1^0$) is greater than internal W-emission process with charged kaonic states (i.e. $K^{*+}$, $K_1^+$) and the difference between the annihilation amplitudes could be ignored~\cite{Cheng:2019ggx}. Take $D_s^+ \to K^{*0}\rho^+$ and $K^{*+}\rho^0$ states as an example, the tree $T$-diagrams and annihilation $A$-diagrams of these two decay modes are shown in Fig.~\ref{fig:topology1} and Fig.~\ref{fig:topology2}, respectively. More experimental information from the amplitude analysis of this decay will offer important experimental input to improve the theory predictions and explore charge-parity ($CP$) violation in the charm meson decays~\cite{PLB684-137,acp2}.

The amplitude analysis of $D_{s}^{+} \to K^+\pi^{+}\pi^{-}\pi^{0}$ also provides access to $D^{+}_s \to VP$ decays, such as $D_s^+ \to \omega K^+$. Evidence for $D_s^+ \to \omega K^+$ was first reported by BESIII experiment, and the BF was measured to be $(0.87\pm0.25_{\rm{stat.}}\pm0.07_{\rm{syst.}})\times 10^{-3}$~\cite{LUYU}, which was based on 3.19 fb$^{-1}$ data samples taken at the center-of-mass energy ($E_{\rm cm}$ or $\sqrt s$) 4.178 GeV. The predicted value of BF (2.12 $\times$ 10$^{-3})$~\cite{Cheng:2019ggx} was too large compared to the experimental value of (0.87 $\times$ 10$^{-3})$, but after taking into account SU(3)$_F$ breaking in internal W-emission, the predicted BF now is reduced to (0.99 $\times$ 10$^{-3}$)~\cite{Cheng:2021yrn}. Therefore, the amplitude of $D_s^+ \to \omega K^+$ decay is important to investigate the W-annihilation contribution in $D^{+}_s \to VP$ decays and improve the understanding of SU(3)$_F$ flavor symmetry breaking effects in hadronic decays of charmed mesons~\cite{PRD93-114010,Cheng:2019ggx,Cheng:2021yrn}.

This paper reports the first amplitude analysis and BF measurement of the decay $D_{s}^{+} \to K^+\pi^{+}\pi^{-}\pi^{0}$, using $e^+e^-$ collision data samples corresponding to an integrated luminosity of 6.32 fb$^{-1}$ collected at the $\sqrt s$ between 4.178 and 4.226 GeV with the BESIII detector. Charged-conjugate modes are always implied throughout this paper except when discussing $CP$ violation.
%------------------------------------------------------------------------------
\section{Detector and data sets}
\label{sec:detector_dataset}
The BESIII detector is a magnetic spectrometer~\cite{Ablikim:2009aa,Ablikim:2019hff} located at the Beijing
Electron Positron Collider (BEPCII)~\cite{Yu:IPAC2016-TUYA01}. The cylindrical core of the BESIII detector
covers 93\% of the full solid angle and consists of a helium-based multilayer
drift chamber~(MDC), a plastic scintillator time-of-flight system~(TOF), and a CsI(Tl)
electromagnetic calorimeter~(EMC), which are all enclosed in a superconducting
solenoidal magnet providing a 1.0~T magnetic field. The solenoid is supported by
an octagonal flux-return yoke with resistive plate counter muon identification
modules interleaved with steel.
%The acceptance of charged particles and photons is 93\% over $4\pi$ solid angle.
The resolution of charged-particle momentum at $1~{\rm GeV}/c$ is $0.5\%$, and the resolution of specific energy
loss $dE/dx$ is $6\%$ for electrons from Bhabha scattering. The EMC
measures photon energies with a resolution of $2.5\%$ ($5\%$) at $1$~GeV in the
barrel (end cap) region. The time resolution in the TOF barrel region is 68~ps,
while that in the end cap region is 110~ps. The end cap TOF
system was upgraded in 2015 using multi-gap resistive plate chamber
technology, providing a time resolution of 60~ps~\cite{etof1, etof2, etof3}.
About 83\% of the data in this analysis benefits from the upgrade.

%------------------------------------------------------------------------------
The integrated luminosities of different centre-of-mass energies of the data samples used in this analysis are listed in Table~\ref{energe}~\cite{XYZLumi,lumin,centermass}. For
some aspects of the analysis, these samples are organised into three sample
groups, 4.178~GeV, 4.189-4.219~GeV, and 4.226~GeV, and each of them is acquired during the
same year under consistent running conditions. Since the cross section of
$D_{s}^{*\pm}D_{s}^{\mp}$ production in $e^{+}e^{-}$ annihilation is about a
factor of twenty larger than that of $D_{s}^{+}D_{s}^{-}$~\cite{DsStrDs}, and
the $D_{s}^{*\pm}$ meson decays to $\gamma D_{s}^{\pm}$ have a dominant BF of
$(93.5\pm0.7)$\%~\cite{PDG}, the signal events discussed in this paper are
selected from the process
$e^+e^-\to D_{s}^{*\pm}D_{s}^{\mp}\to \gamma D_{s}^{+}D_{s}^{-}$.

 \begin{table}[t]
\setlength{\abovecaptionskip}{0.4cm}
\setlength{\belowcaptionskip}{0.2cm}
 \renewcommand\arraystretch{1.25}
 \centering
 \begin{tabular}{|ccc|}
 \hline
 $\sqrt{s}$ (GeV) & $\mathcal{L}_{\rm int}$ (pb$^{-1}$) &\\
 \hline
  4.178 &3189.0$\pm$0.2$\pm$31.9 &\\
  4.189 & 526.7$\pm$0.1$\pm$ 2.2 &\\
  4.199 & 526.0$\pm$0.1$\pm$ 2.1 &\\
  4.209 & 517.1$\pm$0.1$\pm$ 1.8 &\\
  4.219 & 514.6$\pm$0.1$\pm$ 1.8 &\\
  4.226 &1056.4$\pm$0.1$\pm$ 7.0 &\\
  \hline
 \end{tabular}
 \caption{The integrated luminosities ($\mathcal{L}_{\rm int}$) for various centre-of-mass energies. The first and second uncertainties are statistical and systematic, respectively.}
 \label{energe}
\end{table}

Simulated samples produced with a {\sc geant4}-based~\cite{GEANT4} Monte
Carlo (MC) package, which includes the geometric description of the BESIII
detector and the detector response, are used to determine detection
efficiencies and to estimate backgrounds. The simulation models the beam energy
spread and initial state radiation in the $e^+e^-$ annihilations with the
generator {\sc kkmc}~\cite{KKMC1, KKMC2}. The inclusive MC sample includes the
production of open charm processes, the initial state radiation production of vector
charmonium(-like) states, and the continuum processes incorporated in
{\sc kkmc}~\cite{KKMC1, KKMC2}. The known decay modes are modeled with
{\sc evtgen}~\cite{EVTGEN1, EVTGEN2} using BFs taken from the Particle Data
Group~(PDG)~\cite{PDG}, and the remaining unknown charmonium decays are modeled with
{\sc lundcharm}~\cite{LUNDCHARM1, LUNDCHARM2}. Final state radiation from
charged final state particles is incorporated using {\sc photos}~\cite{PHOTOS}.
%------------------------------------------------------------------------------
\section{Event selection}
\label{ST-selection}
%The data samples were collected just above the $D_s^{*\pm}D_s^{\mp}$ threshold.
To obtain the signal samples with high purity, we adopt the double tag method~\cite{DTmethod} in this analysis. In this method, a single-tag~(ST)
candidate requires only one of the $D_{s}^{\pm}$ mesons to be reconstructed via
a hadronic decay; a double-tag~(DT) candidate has both $D_s^+D_s^-$ mesons
reconstructed via hadronic decays,
%The DT candidates are required to have the $D_{s}^{+}$ meson decaying to the signal mode $D_{s}^{+} \to K^+\pi^{+}\pi^{-}\pi^{0}$ and the $D_{s}^{-}$ meson decaying to a tag mode.
where one $D_s$ meson is reconstructed via the signal mode and the other via any of the tag modes.
%In consideration of large statistics and low background contamination, six tag modes are reconstructed and combined to analyze the signal side of $D_{s}^{+} \to K^+\pi^{+}\pi^{-}\pi^{0}$.
%The corresponding mass windows on the tagged $D_{s}^{-}$ mass~($M_{\rm tag}$) are listed in Table~\ref{tab:tag-cut}.
The $D_{s}^{\pm}$ candidates are constructed from
individual $\pi^\pm$, $\pi^{0}$, $K^\pm$, $K_{S}^{0}$, $\eta$ and $\eta^{\prime}$ particles, with the following selection criteria.

All charged tracks reconstructed in the MDC must satisfy $|\!\cos\theta|< 0.93$, where $\theta$ is
the polar angle of a charged track with respect to the positive direction of the MDC axis.
For charged tracks not originating from $K_S^0$ decays, the distance of closest
approach to the interaction point is required to be less than 10~cm along the
beam direction and less than 1~cm in the plane perpendicular to the beam.
Particle identification~(PID) for charged tracks is performed by using the $dE/dx$
measured by the MDC and the flight time in the TOF.
The confidence level for pion and kaon hypotheses~($CL_K$ and $CL_{\pi})$ are calculated. Kaon and pion candidates are required to satisfy $CL_K > CL_{\pi}$ and
$CL_{\pi} > CL_K$, respectively.
%------------------------------------------------------------------------------

The $K_{S}^0$ candidates are reconstructed from two oppositely charged tracks. The
distances of the charged tracks to the interaction point along the beam direction
are required to be less than 20~cm. The two charged tracks are assigned
as $\pi^+\pi^-$ without imposing further PID criteria. They are constrained to
originate from a common vertex and are required to have an invariant mass in the interval of $|M_{\pi^{+}\pi^{-}} - m_{K_{S}^{0}}|<$ 12~MeV$/c^{2}$, where
$m_{K_{S}^{0}}$ is the known $K^0$ mass~\cite{PDG}. The
decay lengths of the $K^0_S$ candidates are required to be twice greater than its uncertainty.

Photon candidates are identified by their showers in the EMC. The deposited
energy of each shower must be more than 25~MeV in the barrel
region~($|\!\cos\theta|< 0.80$) and more than 50~MeV in the end cap
region~($0.86 <|\!\cos\theta|< 0.92$). The minimum opening angle between the position of each
shower in the EMC and the closest extrapolated charged track is required to be greater
than $10^\circ$ to exclude the showers originating from tracks. The
difference between the EMC time and the event start time is required to be
within [0, 700]\,ns to suppress electronic noises and showers unrelated to the
event.

%------------------------------------------------------------------------------
The $\pi^0$ and $\eta$ candidates are formed from the photon pairs with
invariant masses being in the ranges $[0.115, 0.150]$~GeV/$c^{2}$ and
$[0.490, 0.580]$~GeV/$c^{2}$, respectively, which are about three times of the
resolution of the detector. Moreover, at least one of this two photons is required
to be from the barrel EMC to achieve better resolution. A kinematic fit that constrains
the $\gamma\gamma$ invariant mass to the known $\pi^{0}$ or $\eta$ mass~\cite{PDG}
is performed to improve the reconstructed $D_s^{\pm}$ mass resolution. The $\chi^2$ of the kinematic fit is required
to be less than 30. The $\eta^{\prime}$ candidates are formed from
$\pi^{+}\pi^{-}\eta$ combinations with an invariant mass within the range of
$[0.946, 0.970]$~GeV/$c^{2}$.

Six tag modes are used and combined to select the signals of $D_{s}^{+} \to K^+\pi^{+}\pi^{-}\pi^{0}$. The corresponding mass windows on the tagged $D_{s}^{-}$ mass~($M_{\rm tag}$) are listed in Table~\ref{tab:tag-cut}.
The quantity $M_{\rm rec}$ is defined as
\begin{eqnarray}
\begin{aligned}
    M_{\rm rec} = \sqrt{\left(E_{\rm cm} - \sqrt{|\vec{p}_{D_{s}}|^{2}+m_{D_{s}}^{2}}\right)^{2} - |\vec{p}_{D_{s}} | ^{2}} \; , \label{eq:mrec}
\end{aligned}\end{eqnarray}
where $\vec{p}_{D_{s}}$ is the three-momentum of the $D_{s}^-$ candidate in the
$e^+e^-$ centre-of-mass frame, and $m_{D_{s}}$ is the known $D_{s}$ mass~\cite{PDG}.
Events with both signal and tag  $D_{s}$ candidates having their $M_{\rm rec}$ falling within
the bounds listed in Table~\ref{tab:mrec} are retained for further study.
%------------------------------------------------------------------------------
\begin{table}[t]
\setlength{\abovecaptionskip}{0.2cm}
\setlength{\belowcaptionskip}{0.cm}
 \renewcommand\arraystretch{1.25}
 \centering
     \begin{tabular}{|lc|}
        \hline
        Tag mode                                     & Mass window (GeV/$c^{2}$) \\
        \hline
		$D^-_s\to K^0_SK^-$                                   &[1.948, 1.991]\\
		$D^-_s\to K^-K^+\pi^-$                                &[1.950, 1.986]\\
		$D^-_s\to K^+K^-\pi^-\pi^0$                           &[1.947, 1.982]\\
		$D^-_s\to K^0_SK^+\pi^-\pi^-$                         &[1.953, 1.983]\\
		$D^-_s\to \pi^-\eta_{\gamma\gamma}$                   &[1.930, 2.000]\\
		$D^-_s\to \pi^-\eta'_{\pi^+\pi^-\eta_{\gamma\gamma}}$ &[1.938, 1.997]\\
        \hline
      \end{tabular}
 \caption{The $M_{\rm tag}$ requirements for various tag modes, where the subscripts of $\eta$ and $\eta'$ denote the decay modes used to reconstruct these particles.}
\label{tab:tag-cut}
\end{table}
%------------------------------------------------------------------------------
 \begin{table}[t]
\setlength{\abovecaptionskip}{0.2cm}
\setlength{\belowcaptionskip}{0.2cm}
 \renewcommand\arraystretch{1.25}
 \centering
 \begin{tabular}{|cc|}
 \hline
 $\sqrt{s}$ (GeV)  & $M_{\rm rec}$ (GeV/$c^2$)\\
 \hline
  4.178 &[2.050, 2.180] \\
  4.189 &[2.048, 2.190] \\
  4.199 &[2.046, 2.200] \\
  4.209 &[2.044, 2.210] \\
  4.219 &[2.042, 2.220] \\
  4.226 &[2.040, 2.220] \\
  \hline
 \end{tabular}
 \caption{The requirements of $M_{\rm rec}$ for each data set.}
 \label{tab:mrec}
\end{table}

\section{Amplitude analysis}
\label{Amplitude-Analysis}
\subsection{Further selection criteria}
\label{sec:AASelection}
To obtain data samples with high purities for the amplitude analysis, the following dedicated selection criteria are imposed on the signal candidates.

The seven-constraint kinematic fit to the process
$e^+e^-\to D_{s}^{*\pm}D_{s}^{\mp}\to \gamma D_{s}^{+}D_{s}^{-}$,
where the $D_{s}^{-}$ decays to one of the tag modes and the $D_{s}^{+}$ decays to
the signal mode, is required to converge. In addition to the constraints of four-momentum
conservation in the $e^+e^-$ centre-of-mass system, the invariant masses of
$(\gamma\gamma)_{\pi^{0}}$, tag $D_{s}^{-}$, and $D_{s}^{*\pm}$ candidates are constrained to the
corresponding known masses~\cite{PDG}. The combination with the minimum $\chi^2$ is chosen, assuming
that $D^{*+}_s$ decays to $D_{s}^{+}\gamma$ or $D_s^{*-}$ decays to $D_{s}^{-}\gamma$.
 In order to ensure that all
candidates fall within the phase-space boundary, the constraint of the signal
$D_s^{+}$ mass is added to the kinematic fit and the updated four-momenta
from this kinematic fit are used for the amplitude analysis.

%------------------------------------------------------------------------------
A $K_S^0$ mass veto, $M_{\pi^+\pi^-} \notin$ [0.460, 0.520]~GeV/$c^2$, which is about $\pm$3 times of the $K_S^0$ resolution, is applied on the signal $D_s^+$ to remove the dominant background from $D_s^+ \to K^+K_S^0\pi^0$ ($K_S^0\to\pi^+\pi^-$) decays. An $\eta$ mass veto, $M_{\pi^+\pi^-\pi^0} \notin$ [0.520, 0.580]~GeV/$c^2$, which is about $\pm$3 times the $\eta$ resolution, is also applied to remove the events from $D_s^+ \to K^+\eta$ ($\eta \to \pi^+\pi^-\pi^0$) decays.

The energy of the transition photon from $D_s^{*+}\to \gamma D_s^{+}$
is required to be less than 0.18~GeV. The recoiling mass against this photon and the signal
$D_s^{+}$ candidate is required to lie within the range of $[1.955, 1.995]$~GeV/$c^2$.

There is a wrong-combination background from $D^0\to K^-\pi^+\pi^0$ versus
$\bar{D}^0\to K^+\pi^+\pi^-\pi^-$, which fakes $D_s^+ \to K^+\pi^+\pi^-\pi^0$ versus
$D_s^- \to K^-K_S^0$, $K_S^0 \to \pi^+\pi^-$ by exchanging a $\pi^0$ and $\pi^-$.
It also fakes  $D_s^+ \to K^+\pi^+\pi^-\pi^0$ versus $D_s^- \to K^+K^-\pi^-$
by identifying a $\pi^+$ from the $D^0$  as a $K^+$
and exchanging a $\pi^-$ from the $\bar{D}^0$ with the $\pi^0$ from $D^0$.
This background is excluded by rejecting the events which simultaneously satisfy
$|M_{K^-\pi^+\pi^0}-M_{D^{0}}|< 75$~MeV/$c^2$ and $|M_{K^+\pi^+\pi^-\pi^-}-M_{\bar{D}^0}|< 50$~MeV/$c^2$, where $M_{D^{0}}$ is the known $D^0$ mass~\cite{PDG}.
There is also a wrong-combination background from $D^+ \to K^+K_S^0K_S^0$, $K_S^0 \to \pi^+\pi^-$ versus $D^- \to K^+\pi^-\pi^-$, it fakes $D_s^+ \to K^+\pi^+\pi^-\pi^0$ versus $D_s^- \to K_S^0K^+\pi^-\pi^-$ by exchanging $\pi^+$ and $\pi^-$, then adding a $\pi^0$. This background is excluded by rejecting the events which simultaneously satisfy $|M_{K^+\pi^+\pi^-\pi^+\pi^-}-M_{D^+}|< 50$~MeV/$c^2$ and $|M_{K^+\pi^-\pi^-}-M_{D^-}|< 75$~MeV/$c^2$, where $M_{D^+}$ is the known $D^+$ mass~\cite{PDG}.

\begin{figure*}[!htbp]
\setlength{\abovecaptionskip}{-0.2cm}
\setlength{\belowcaptionskip}{0.cm}
  \centering
  \includegraphics[width=0.495\textwidth]{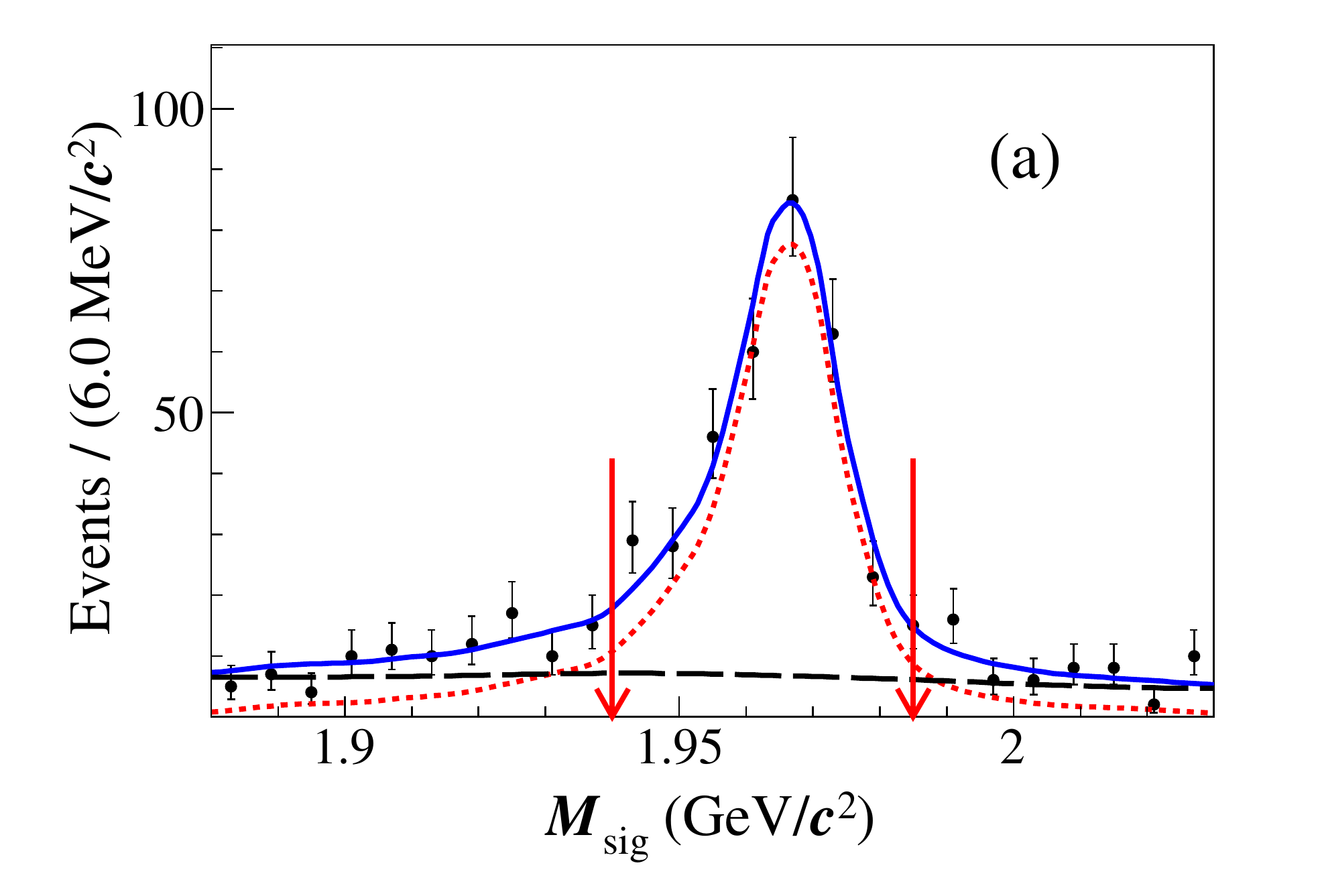}
  \includegraphics[width=0.495\textwidth]{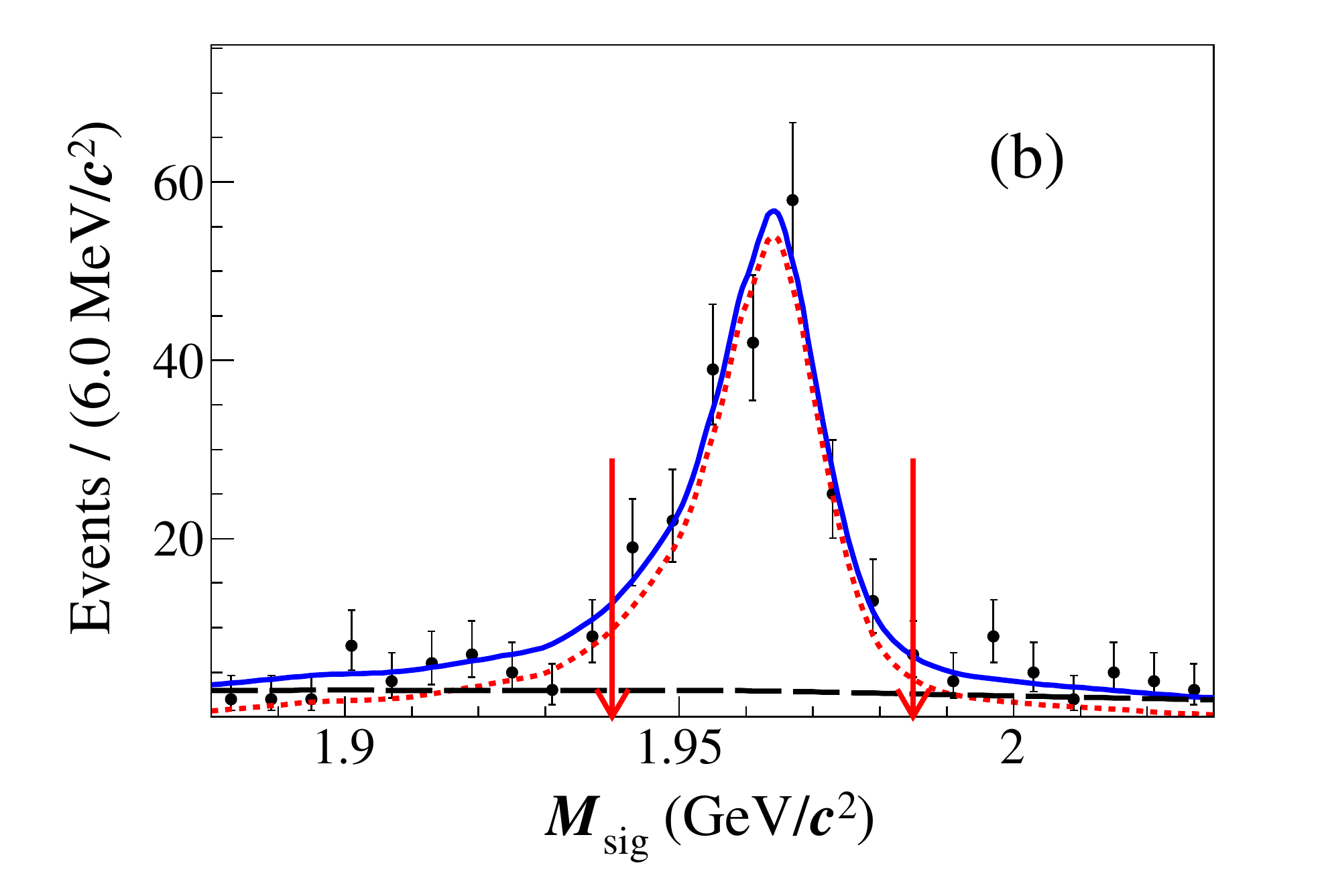}
  \includegraphics[width=0.495\textwidth]{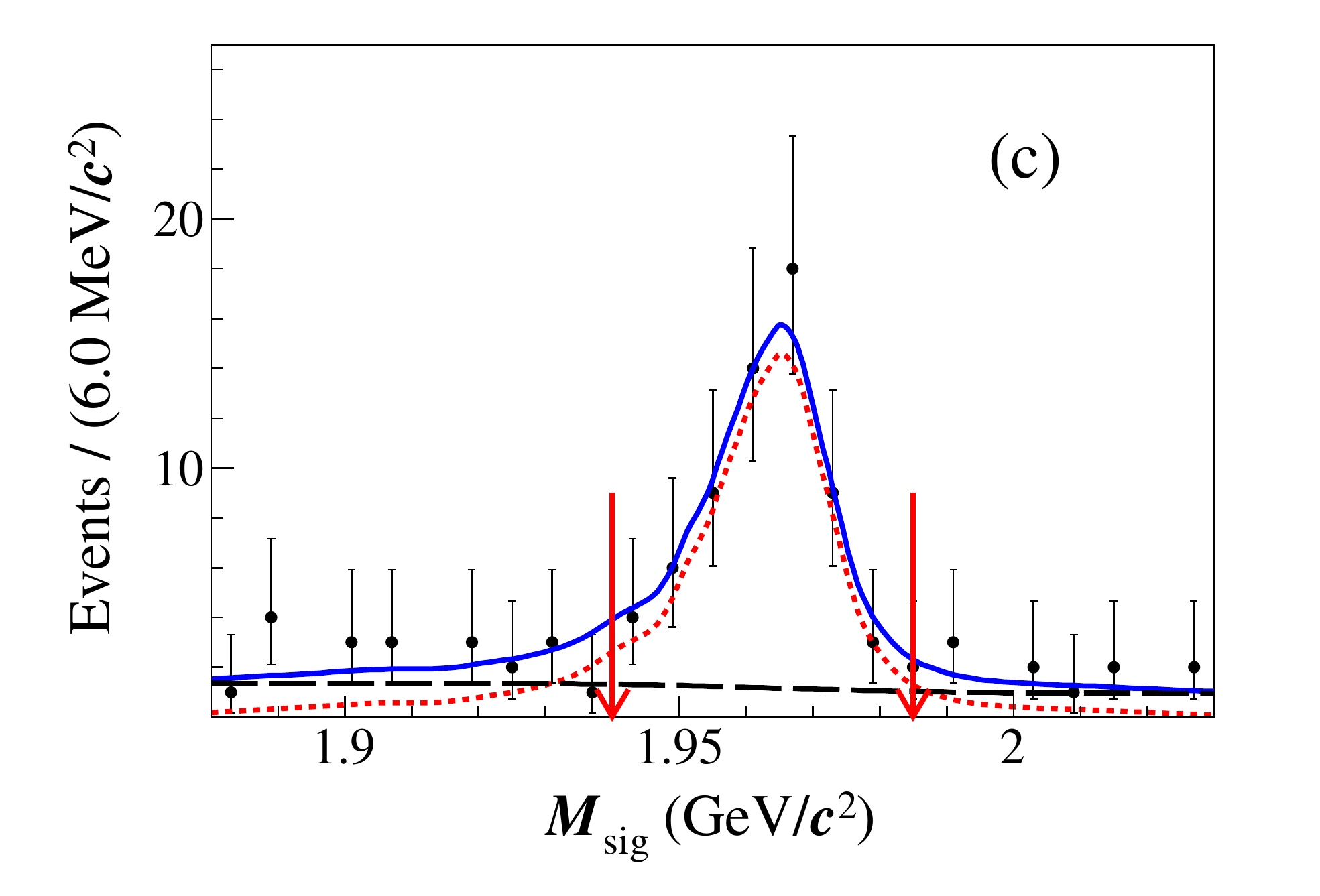}
  \includegraphics[width=0.495\textwidth]{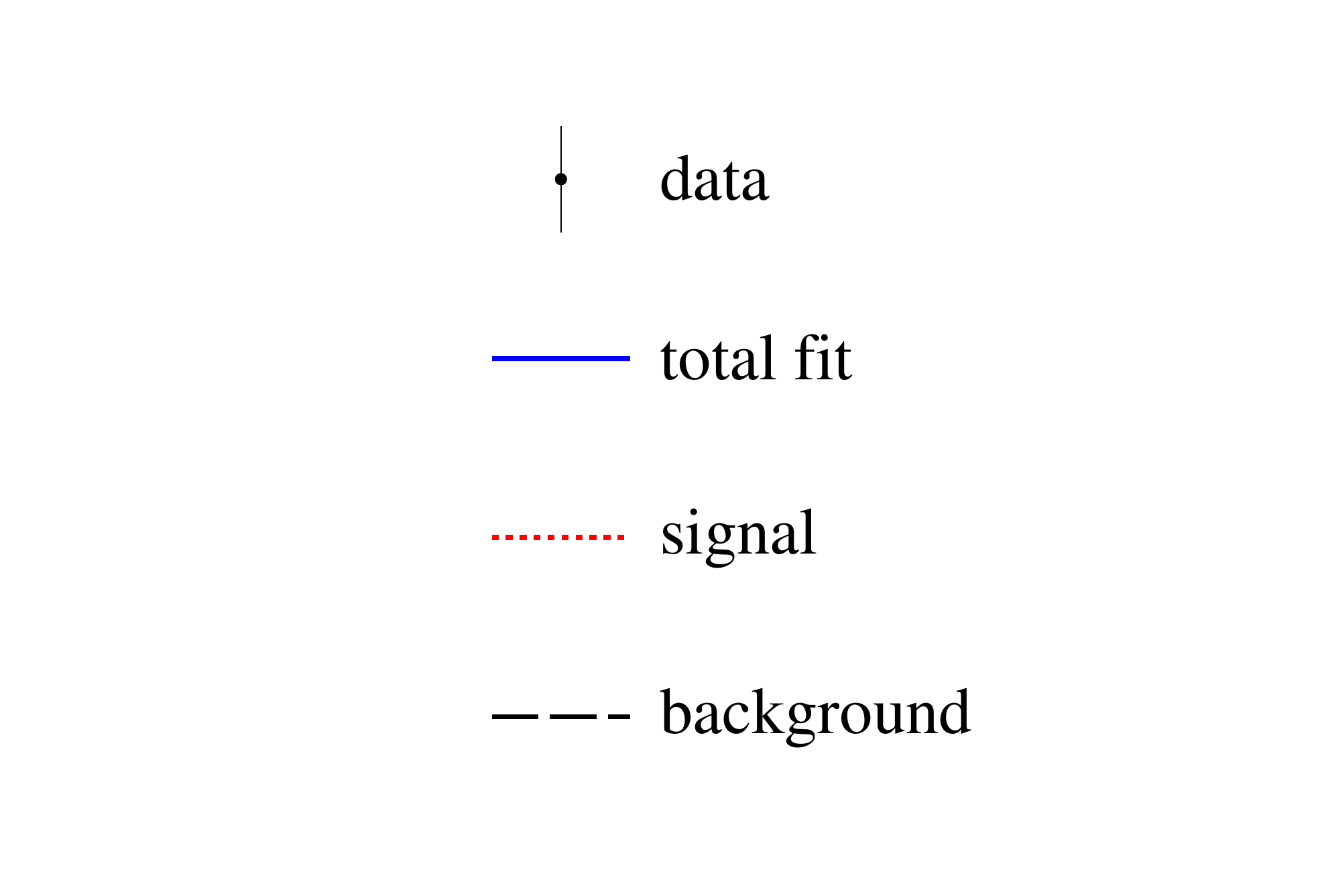}
  \caption{
    Fits to the $M_{\rm sig}$ distributions of the data samples at $\sqrt{s}=$
    (a) 4.178~GeV, (b) 4.189-4.219~GeV and (c) 4.226~GeV. The black points with
    error bars are data. The blue solid lines are the total fits. The red dotted
    and black dashed lines are the fitted signal and background, respectively.
    The pairs of red arrows indicate the signal regions.
  } \label{fig:fit_Ds}
\end{figure*}

Figure~\ref{fig:fit_Ds} shows the
fits to the invariant-mass distributions of the accepted signal $D_s^{+}$
candidates, $M_{\rm sig}$, for various data samples. In the fits, the signal is described by a MC-simulated shape convolved with a Gaussian function and the background is described by a simulated shape derived from the inclusive MC sample.
Then, a mass window, $[1.940, 1.985]$~GeV/$c^2$, is applied on the signal $D_s^{+}$ candidates.
Finally, there are 344, 222, and 64 events retained for the amplitude analysis
with purities of $(85.1\pm1.9)\%$, $(90.0\pm2.0)\%$, and $(86.1\pm4.3)\%$ for the
data samples at $\sqrt{s} = 4.178$, 4.189-4.219, and 4.226 GeV, respectively.

\subsection{Fit method}
\label{sec:fitmethod}
%The intermediate-resonant composition is determined by an unbinned maximum-likelihood fit to data.
An unbinned maximum likelihood fit is used in the amplitude analysis of $D_s^+ \to K^+\pi^+\pi^-\pi^0$, the likelihood function $\mathcal{L}$ is constructed with a signal-background combined probability density
function~(PDF), which depends on the momenta of the four final state particles.
The likelihood is written as
\begin{eqnarray}\begin{aligned}
  \mathcal{L} = \prod_{i=1}^{3}\prod_{k=1}^{N_{D, i}}\left[w^{i}f_{S}(p^{k})+(1-w^{i})f_{B}(p^k)\right]\,,  \label{likelihood3}
\end{aligned}\end{eqnarray}
where $i$ indicates the data sample groups. The $p^k$ denotes the four-momenta of the final state particles, where $k$ denotes the $k^{\rm th}$ event in the data sample $i$. The $N_{D,i}$ is the number of candidates in the data sample $i$,
$f_{S}$~($f_{B}$) is the signal~(background) PDF and $w^i$ is the purity of the signal discussed in Sec.~\ref{sec:AASelection}.

The signal PDF is given by
\begin{eqnarray}\begin{aligned}
  f_{S}(p) = \frac{\epsilon(p)\left|\mathcal{M}(p)\right|^{2}R_{4}}{\int \epsilon(p)\left|\mathcal{M}(p)\right|^{2}R_{4}\,dp}\,, \label{signal-PDF}
\end{aligned}\end{eqnarray}
where $\epsilon(p)$ is the detection efficiency in bins of a five-dimensional space of two- and three-body invariant masses, and $R_{4}$ is the four-body phase space. %The isobar formalism is used to model the total amplitude. Constructed with covariant tensors,
The total amplitude $\mathcal{M}$ is modeled with the isobar model, which is the coherent sum of the individual amplitudes of intermediate processes,
given by $\mathcal{M}=\sum \rho_{n}e^{i\phi_{n}}\mathcal{A}_{n}$, where the magnitude $\rho_{n}$ and
phase $\phi_{n}$ are the free parameters to be determined by the fit. The amplitude of the $n^{\rm th}$ intermediate process~($\mathcal{A}_{n}$)
is given by
%The magnitude $\rho_{n}$ and phase $\phi_{n}$ are free parameters in the fit,
%and are defined relative to those of a reference mode, for which they are fixed.
\begin{eqnarray}
\begin{aligned}
  \mathcal{A}_{n} = P_n^1P_n^2S_nF_n^1F_n^2F_n^{3}\,, \label{base-amplitude}
\end{aligned}\end{eqnarray}
where the indices 1, 2 and 3 correspond to the two subsequent intermediate resonances and the $D^+_s$ meson. Here
$F_{n}$ is the Blatt-Weisskopf barrier (Sec.~\ref{sec:barrier}),
$P_{n}$ is the propagator of the intermediate resonance (Sec.~\ref{sec:propagator}), and
$S_{n}$ is the spin factor constructed with the covariant tensor formalism~\cite{covariant-tensors} (Sec.~\ref{sec:spinfactor}).
The normalisation integral is realised by MC integration,
\begin{eqnarray}\begin{aligned}
  \int \epsilon(p) |\mathcal{M}(p)|^2 R_{4}\,dp \approx
\frac{1}{N_{\rm MC}}\sum_{k=1}^{N_{\rm MC}} \frac{ |\mathcal{M}(p^{k})|^2 }{\left|\mathcal{M}^{g}(p^{k})\right|^{2}}\,, \label{MC-intergral}
\end{aligned}\end{eqnarray}
where $k$ is the index of the $k^{\rm th}$ event of the signal MC sample, and $N_{\rm MC}$ is the number of the selected MC events.
The $\mathcal{M}^{g}(p)$ is the signal PDF used to generate the signal MC sample in the MC integration.
The normalization integral for background is also realised by a MC integration method like Eq. ~\ref{MC-intergral},
\begin{equation}
\int \epsilon(p)B_\epsilon(p)R_4dp \approx \frac{1}{N_{MC}}\sum^{N_{MC}}_{k_{MC}}\frac{B_\epsilon(p^{k_{MC}})}{|M^{gen}(p^{k_{MC}})|^2},
\end{equation}
%These MC samples are generated with different $\sqrt{s}$
%according to the luminosities and cross sections, and satisfy all selection
%criteria as those of the data samples. At the beginning, a
%preliminary PDF is used, and then a recursive process is performed until the
%result converges.

To account for the bias caused by differences in tracking, PID efficiencies and $\pi^0$ reconstruction
    between data and MC simulation, each signal MC event is weighted with a ratio, $\gamma_{\epsilon}(p)$, and it is calculated as
\begin{equation}
	\gamma_{\epsilon}(p) = \prod_{j} \frac{\epsilon_{j,\rm data}(p)}{\epsilon_{j,\rm MC}(p)},
	\label{pwa:gamma}
\end{equation}
where $j$ denotes the final four daughter particles, $\epsilon_{j,\rm data}(p)$ and $\epsilon_{j,\rm MC}(p)$ are the tracking, PID and $\pi^0$ reconstruction efficiencies as a function of the momenta of the daughter particles for data and MC simulation, respectively.
By weighting each signal MC event with $\gamma_{\epsilon}$, the MC integration is modified to be
\begin{eqnarray}\begin{aligned}
    &\int \epsilon(p) |\mathcal{M}(p)|^2 R_{4}\,dp \approx
&\frac{1}{N_{\rm MC}} \sum_{k=1}^{N_{\rm MC}} \frac{ |\mathcal{M}(p^{k})|^2 \gamma_{\epsilon}(p^{k_{\rm MC}})}{\left|\mathcal{M}^{g}(p^{k})\right|^{2}}\,.
\label{MC-intergral-corrected}
\end{aligned}\end{eqnarray}
The background PDF is given by
\begin{eqnarray}\begin{aligned}
  f_{B}(p) = \frac{\epsilon(p)B_{\epsilon}(p)R_{4}}{\int \epsilon(p)B_{\epsilon}(p)R_{4}\,dp}\,,\label{bkg-PDF}
\end{aligned}\end{eqnarray}
where $B_{\epsilon}(p)=B(p)/\epsilon(p)$ is the efficiency-corrected
background shape. The background shape $B(p)$ is derived by using a multi-dimensional kernel
density estimator~\cite{Cranmer} named RooNDKeysPdf implemented in RooFit~\cite{Verkerke},
which models the distribution of an input dataset as a superposition of Gaussian kernels
using background events in the $M_{\rm sig}$ signal region from the inclusive MC sample.
The $M_{K^+\pi^-}$, $M_{K^+\pi^0}$, $M_{\pi^+\pi^-}$, $M_{\pi^+\pi^0}$ and $M_{K^+\pi^-\pi^0}$
distributions of the inclusive MC events outside the $M_{\rm sig}$ signal region are compared to these distributions from the data to check their validity. The distributions of background events from the inclusive MC sample within and
outside the $M_{\rm sig}$ signal region are also
examined. They are compatible with each other within statistical uncertainties.

\subsubsection{Blatt-Weisskopf barrier factors}\label{sec:barrier}
For the process $a \to bc$, the Blatt-Weisskopf barrier factors~\cite{BW},
$X_L(p)$, are
parameterised as a function of the angular momenta $L$ and the momenta $q$ of
the final state particle $b$ or $c$ in the rest system of $a$. They are taken as
\begin{eqnarray}
\begin{aligned}
 X_{L=0}(q)&=1,\\
 X_{L=1}(q)&=\sqrt{\frac{z_0^2+1}{z^2+1}},\\
 X_{L=2}(q)&=\sqrt{\frac{z_0^4+3z_0^2+9}{z^4+3z^2+9}}\,,
\end{aligned}
\end{eqnarray}
where $z=qR$, $z_0=q_0R$ and the effective radius of the barrier $R$ is fixed to 3.0~GeV$^{-1}$ for the intermediate
resonances and 5.0~GeV$^{-1}$ for the $D_s^+$ meson.
The momentum $q$ is given by
\begin{eqnarray}
\begin{aligned}
q = \sqrt{\frac{(s_a+s_b-s_c)^2}{4s_a}-s_b}\,, \label{q2}
\end{aligned}
\end{eqnarray}
the value of $q_0$ is that of $q$ when $s_a = m_a^2$ and the $s_a(s_b,s_c)$ denotes the invariant-mass squared of the particle $a(b,c)$.
%------------------------------------------------------------------------------
\subsubsection{Propagator}\label{sec:propagator}
The intermediate resonances $K^{*0}$, $K^{*+}$, $K_1(1270)^0$, $K_1(1400)^0$ and $a_1(1260)^0$ are
parameterised with the relativistic Breit-Wigner~(RBW) function,
\begin{eqnarray}\begin{aligned}
    P(m) = \frac{1}{m_{0}^{2} - s_{a} - im_{0}\Gamma(m)}\,,\;
    \Gamma(m) = \Gamma_{0}\left(\frac{q}{q_{0}}\right)^{2L+1}\left(\frac{m_{0}}{m}\right)\left(\frac{X_{L}(q)}{X_{L}(q_{0})}\right)^{2}\,,
  \label{RBW}
\end{aligned}\end{eqnarray}
where $m_0$ and $\Gamma_0$ denote the resonance's rest mass and width. The masses and widths of the intermediate resonances, except for $K_1(1270)^0$, are fixed to the PDG values~\cite{PDG}. Considering the obvious mass deviation reported in the PDG~\cite{PDG}, the mass and width of $K_1(1270)^0$ are fixed to 1289 MeV/$c^2$ and 116 MeV, respectively, from results obtained by the LHCb experiment~\cite{K1270}.
%------------------------------------------------------------------------------

The $\rho$ resonances are parameterised by the
Gounaris-Sakurai~(GS) lineshape~\cite{GS}, which is given by
\begin{eqnarray}\begin{aligned}
P_{\rm GS}(m)=\frac{1+d\frac{\Gamma_0}{m_0}}{m_0^2-m^2+f(m)-im_0\Gamma(m)}\,.
\end{aligned}\end{eqnarray}
The function $f(m)$ is given by
\begin{eqnarray}
\begin{aligned}
f(m)=\Gamma_0\frac{m_0^2}{q_0^3}\left[q^2(h(m)-h(m_0))+(m_0^2-m^2)q_0^2\left.\frac{dh}{d(m^2)}\right|_{m_0^2}\right]\,,
\end{aligned}
\end{eqnarray}
where
\begin{eqnarray}\begin{aligned}
h(m)=\frac{2q}{\pi m}\ln\left(\frac{m+2q}{2m_{\pi}}\right)\,,
\end{aligned}\end{eqnarray}
and
\begin{eqnarray}\begin{aligned}
&\left.\frac{dh}{d(m^2)}\right|_{m_0^2}=
&h(m_0)\left[(8q_0^2)^{-1}-(2m_0^2)^{-1}\right]+(2\pi m_0^2)^{-1}\,.
\end{aligned}\end{eqnarray}
The normalisation condition at $P_{\rm GS}(0)$ fixes the parameter
$d=f(0)/(\Gamma_0 m_0)$ as
\begin{eqnarray}\begin{aligned}
d=\frac{3m^2_\pi}{\pi q_0^2}\ln\left(\frac{m_0+2q_0}{2m_\pi}\right)+\frac{m_0}{2\pi q_0}-\frac{m^2_\pi m_0}{\pi q^3_0}\,.
\end{aligned}\end{eqnarray}

The $K$-Matrix parametrisation is used to describe the $\pi^+\pi^-$ S-wave. Detailed descriptions of the $K$-matrix formalism can be found in various references~\cite{PPs1,PPs2,PPs3,PPs4}; parameters used are summarised in Tables~\ref{tab:pipi_para1} and~\ref{tab:pipi_para2}. We use the ``$K$-matrix amplitude'' to describe the amplitude of channel $u$ ($u = 1-5$ denote the channels $\pi\pi, K\bar{K}, 4\pi, \eta\eta, \eta\eta'$) in the form of $A_u=(I-i\hat{K}\rho)^{-1}_{uv}\hat{P}_v$. Here the vector $\hat{P}$ describes the production of bare states and the non-resonant production of meson pairs, while the term $(I-i\hat{K}\rho)^{-1}$ describes their re-scattering.

\begin{table}[htbp]
        \caption{K-matrix parameters from a global analysis of the available $\pi\pi$ scattering data from threshold up to 1900 MeV/$c^2$. Masses and coupling constants are given in GeV/$c^2$}
        \centering
        \begin{tabular}{lcccccc}
        \hline
        \hline
          $\alpha$ &$m_{\alpha}$  &$g_{\pi^+\pi^-}^{\alpha}$  &$g_{K\bar{K}}^{\alpha}$  &$g_{4\pi}^{\alpha}$ &$g_{\eta\eta}^{\alpha}$ &$g_{\eta\eta}^{\alpha}$\\
          \hline
        1 &0.65100      &0.22889      &$-0.55377$    &0.00000      &$-0.398994$  &$-0.34639$\\
        2 &1.20360      &0.94128      &0.55095       &0.00000      &0.39065      &0.31503\\
        3 &1.55817      &0.36856      &0.23888       &0.55639      &0.18340      &0.18681\\
        4 &1.21000      &0.33650      &0.40907       &0.85679      &0.19906      &$-0.00984$\\
        5 &1.82206      &0.18171      &$-0.17558$    &$-0.79658$   &$-0.00355$   &0.22358\\
        \hline
        $s_{0}^{\rm sacct}$     &$f_{11}^{\rm sacct}$     &$f_{12}^{\rm sacct}$     &$f_{13}^{\rm sacct}$      &$f_{14}^{\rm sacct}$      &$f_{15}^{\rm sacct}$\\
        $-3.92637$      &0.23399      &0.15044      &$-0.20545$      &0.32825 &0.35412\\
        \hline
        $s_{A0}$   &$s_A$   &$s_{0}^{\rm prod}$        \\
        -0.15      &1.0     &-3.0 $\pm$ 0.03        \\
        \hline
        \hline
        \end{tabular}
        \label{tab:pipi_para1}
\end{table}
\begin{table}[htbp]
        \caption{$\pi\pi$ S-wave P-vector parameters obtained from the $D^0 \to K_S^0\pi^+\pi^-$ Dalitz plot distribution from $D^{*+} \to D^0\pi^+$. P-vector parameters $f_{1v}^{'prod}$, for $v \neq 1$, are defined as $f_{1v}^{prod}$/$f_{11}^{prod}$.}
        \centering
        \begin{tabular}{l c@{ $\pm$ }c c@{ $\pm$ }c}
        \hline
        \hline
          Component  &\multicolumn{2}{c}{$a_r$}  &\multicolumn{2}{c}{$\phi_r$(deg)}\\
        \hline
   $\beta_1$  &9.3   &0.4   &$-78.7$  &1.6 \\
   $\beta_2$  &10.89 &0.26  &$-159.1$ &2.6 \\
   $\beta_3$  &24.2  &2.0   &168.0  &4.0 \\
   $\beta_4$  &9.16  &0.24  &90.5   &2.6 \\
   $f_{11}^{prod}$   &7.94   &0.26   &73.9  &1.1\\
   $f_{12}^{'prod}$  &2.0    &0.3    &$-18.0$  &9\\
   $f_{13}^{'prod}$  &5.1    &0.3    &33  &3 \\
   $f_{14}^{'prod}$  &3.23   &0.18   &4.8  &2.5\\
        \hline
        \hline
        \end{tabular}
        \label{tab:pipi_para2}
\end{table}

The scattering matrix $\hat{K}$ can be parameterised as a combination of the sum of $N$ poles
with real bare masses $m_\alpha$, together with slowly-varying non-resonant parts (SVPs):
\begin{equation}
    \hat{K}_{uv}(s)=\left(\sum\limits_{\alpha=1}^{N_{\rm poles}}\frac{g_u^{\alpha}g_v^{\alpha}}{m_{\alpha}^2-s}+f_{uv}^{\rm scatt}\frac{m_0^2-s_{0}^{\rm scatt}}{s-s_{0}^{\rm scatt}}\right)\left[\frac{1\ {\rm GeV}^2/c^4-s_{A_0}}{s-s_{A_0}}(s-s_{A}m_{\pi}^2/2)\right],
\end{equation}
where $g_u^{\alpha}$ denotes the real coupling constant of the pole $m_{\alpha}$ to the meson channel $u$. The parameters $f_{uv}^{\rm scatt}$ and $s_{0}^{\rm scatt}$ describe a smooth part for the $K$-matrix elements and $m_0^2$, $s_A$, and $s_{A_0}$ are real constants of order unity. All these parameters are taken from Ref.~\cite{PPs2}. Here $s$ denotes the invariant mass squared of $\pi^+\pi^-$.

The production vector $\hat{P}$ vector is parameterised in a form analogous to the $\hat{K}$ matrix and it is given by
\begin{equation}
\label{pipis}
\hat{P}_v(s)=f_{v}^{\rm prod}\frac{1-s_{0}^{\rm prod}}{s-s_{0}^{\rm prod}}+\sum\limits_{\alpha}\frac{\beta_{\alpha}g_v^{\alpha}}{m_{\alpha}^2-s},
\end{equation}
where $\beta_{\alpha}$ and $f_{v}^{\rm prod}$ are complex production constants for the poles and non-resonant SVPs, respectively, both of them
depend on the final state channel.

The $K\pi$ S-wave is modeled by a parameterisation from scattering data~\cite{KPsnew}, which is described by a $K_0^{*}(1430)$ Breit-Wigner together with an effective range non-resonant component with a phase shift. It is given by	
\begin{equation}
A(m)=F\sin\delta_Fe^{i\delta_F}+R\sin\delta_Re^{i\delta_R}e^{i2\delta_F},
\end{equation}
with
\begin{equation}
\begin{aligned}
&\delta_F=\phi_F+\cot^{-1}\left[\frac{1}{aq}+\frac{rq}{2}\right], \\
&\delta_R=\phi_R+\tan^{-1}\left[\frac{M\Gamma(m_{K\pi})}{M^2-m^2_{K\pi}}\right],
\end{aligned}
\end{equation}
where the parameters $F(\phi_F)$ and $R(\phi_R)$ are the magnitude (phase) for non-resonant state and resonance terms, respectively. The parameters $a$ and $r$ are the scattering length and effective interaction length, respectively. We fix these parameters $(M, \Gamma, F, \phi_F , R, \phi_R, a, r)$ to the results obtained from the amplitude analysis to a sample of $D^0\to K_S^0\pi^+\pi^-$ by the BABAR and Belle experiments~\cite{KPsnew}; these parameters are summarised in Table~\ref{tab:babar}.
% Parametrisation with the scattering $K$-matrix amplitude from Ref.~\cite{Kx} was also tried and
% gave very close results, the differences are assigned as the associated systematic uncertainties.
\begin{table}[t]
\setlength{\abovecaptionskip}{0.cm}
\setlength{\belowcaptionskip}{-0.2cm}
  \begin{center}
    \begin{tabular}{|lc|}
			\hline
      $M$~(GeV/$c^2$)      & $1.441   \pm 0.002$\\
      $\Gamma$~(GeV)       & $0.193   \pm 0.004$\\
      $F$                  & $0.96    \pm 0.07$ \\
      $\phi_F$~($^\circ$)  & $0.1     \pm 0.3$  \\
      $R$                  & 1(fixed)\\
      $\phi_R$~($^\circ$)  & $-109.7  \pm 2.6$  \\
      $a$~(GeV/$c$)$^{-1}$ & $0.113   \pm 0.006$\\
      $r$~(GeV/$c$)$^{-1}$ & $-33.8   \pm 1.8$  \\
      \hline
    \end{tabular}
  \end{center}
  \caption{The $K\pi$ S-wave parameters, obtained from the amplitude analysis of $D^0\to K_S^0\pi^+\pi^-$ by the BABAR and Belle experiments~\cite{KPsnew}. The uncertainties are combined from the statistical and systematic uncertainties.}
  \label{tab:babar}
\end{table}
%------------------------------------------------------------------------------
\subsubsection{Spin factors}\label{sec:spinfactor}
For the process $a \to bc$, the four-momenta of the particles $a$, $b$, and $c$ are denoted as $p_a$, $p_b$, and $p_c$, respectively.
The spin projection operators~\cite{covariant-tensors} are defined as
\begin{eqnarray}
\begin{aligned}
  P^{(0)}(a) &= 1\,,\\
  P^{(1)}_{\mu\mu^{\prime}}(a) &= -g_{\mu\mu^{\prime}}+\frac{p_{a,\mu}p_{a,\mu^{\prime}}}{p_{a}^{2}}\,,\\
  P^{(2)}_{\mu\nu\mu^{\prime}\nu^{\prime}}(a) &= \frac{1}{2}\left(P^{(1)}_{\mu\mu^{\prime}}(a)P^{(1)}_{\nu\nu^{\prime}}(a)+P^{(1)}_{\mu\nu^{\prime}}(a)P^{(1)}_{\nu\mu^{\prime}}(a)\right)\\
  &-\frac{1}{3}P^{(1)}_{\mu\nu}(a)P^{(1)}_{\mu^{\prime}\nu^{\prime}}(a)\,.
 \label{spin-projection-operators}
\end{aligned}
\end{eqnarray}
The pure orbital angular-momentum covariant tensors are given by
\begin{eqnarray}
\begin{aligned}
    \tilde{t}^{(0)}_{\mu}(a) &= 1\,,\\
    \tilde{t}^{(1)}_{\mu}(a) &= -P^{(1)}_{\mu\mu^{\prime}}(a)r^{\mu^{\prime}}_{a}\,,\\
    \tilde{t}^{(2)}_{\mu\nu}(a) &= P^{(2)}_{\mu\nu\mu^{\prime}\nu^{\prime}}(a)r^{\mu{\prime}}_{a}r^{\nu^{\prime}}_{a}\,,\\
\label{covariant-tensors}
\end{aligned}
\end{eqnarray}
where $r_a = p_b-p_c$. The spin factors $S(p)$ used are listed in Table~\ref{table:spin_factors}. The tensor describing the $D_s^+$ decays with orbital angular-momentum quantum number $l$ is denoted by $\tilde{T}^{(l)\mu}$ and that of intermediate $a \to bc$ decay is denoted by $\tilde{t}^{(l)\mu}$, and the $\tilde{T}^{(l)\mu}$ has the same definition as $\tilde{t}^{(l)\mu}$ in Ref.~\cite{covariant-tensors}.
\begin{table}[t]
\setlength{\abovecaptionskip}{0.cm}
\setlength{\belowcaptionskip}{0.cm}
 \begin{center}
\begin{tabular}{|lc|}
\hline
 Decay chain& $S(p)$ \\
 \hline
$D_s^+[S] \rightarrow V_1V_2$ & $\tilde{t}^{(1)\mu}(V_1) \; \tilde{t}^{(1)}_\mu(V_2)$ \\
$D_s^+[P] \rightarrow V_1V_2$ & $\epsilon_{\mu\nu\lambda\sigma}p^\mu(D_s^+) \; \tilde{T}^{(1)\nu}(D_s^+)\tilde{t}^{(1)\lambda}(V_1) \; \tilde{t}^{(1)\sigma}(V_2) $ \\
$D_s^+ \rightarrow AP_1, A[S] \rightarrow VP_2$ & $\tilde{T}^{(1)\mu}(D_s^+) \; P^{(1)}_{\mu\nu}(A) \; \tilde{t}^{(1)\nu}{(V)}$ \\
$D_s^+ \rightarrow V_1P_1, V_1 \rightarrow V_2P_2$ & $\epsilon_{\mu\nu\lambda\sigma}p^\mu_{V1}r^\nu_{V1}p^\lambda_{P1}r^\sigma_{V2}$ \\
$D_s^+ \rightarrow SS$ & 1 \\
\hline
\end{tabular}
\end{center}
 \caption{The spin factors $S(p)$ for various components. All operators, i.e.~$\tilde{t}$ and ~$\tilde{T}$, have the same definitions as in Ref.~\cite{covariant-tensors}.
  Scalar, pseudo-scalar, vector and axial-vector states are denoted
  by $S$, $P$, $V$ and $A$, respectively. The $[S]$ and $[P]$ denote the orbital angular-momentum quantum numbers $L$ = 0 and 1, respectively.}\label{table:spin_factors}
\end{table}

%The masses and widths of most resonances are fixed to the corresponding PDG averages~\cite{PDG}.
\subsection{Fit results}
Using the method described in Sec.~\ref{sec:fitmethod}, we perform an unbinned maximum likelihood fit to the
$D_s^+ \to K^+ \pi^+\pi^- \pi^0$ decay channel. The fit is performed in steps,
by adding resonances one by one. The corresponding statistical significance for the newly added amplitude is
calculated with the change of the log likelihood value, taking the change of the number of the degrees of freedom into account.

% The fit to the data is initially performed with a baseline model including the amplitude of the
% decay, as it is clearly observed in the invariant mass distributions and tends to be
% the dominant intermediate process.
For the amplitude fits, the magnitude and phase of the $D_s^{+}\to K^{*0}\rho^+$
reference amplitude are fixed to 1 and 0, respectively, while those of the other amplitudes
are floated in the fit.
The amplitudes for $D_s^{+}\to K^{*+}\rho^0$ and $D_s^{+}\to K^+\omega$ are also included,
as they are clearly observed in the corresponding invariant mass spectra.
After testing each, $D_s^{+}\to K_1(1270)^0\pi^+\ (K_1(1270)^0\to K^+ \rho^-)$, $D_s^{+}\to K_1(1400)^0\pi^+\ (K_1(1400)^0\to K^{*\pm}\pi^{\mp})$, $D_s^{+}\to K^+a_1(1260)^0\ (a_1(1260)^0\to \rho^{\pm}\pi^{\mp})$, $D_s^{+}\to (K^+\pi^0)_{V}\rho^0$, and $D_s^{+}\to (K^+\pi^0)_{\rm S-wave}(\pi^+\pi^-)_{\rm S-wave}$ are added since each has a statistical significance greater than $3\sigma$.
%where $(K^+\pi^0)_{\rm S}$ and $(\pi^+\pi^-)_{\rm S}$ denotes $(K^+\pi^0)_{\rm S-wave}$ and $(\pi^+\pi^-)_{\rm S-wave}$ respectively.
Considering the isospin relationship in hadron decays, some Clebsch-Gordan relations are fixed, with details in Appendix~\ref{CGrelation}. A full list of other allowed contributions (based on known states) with statistical significances less than $3\sigma$ are listed in Appendix~\ref{tested_amplitude}.

The fit fraction~(FF) for the $n^{\rm{th}}$ amplitude is computed numerically with generator-level MC events with the definition as
\begin{eqnarray}\begin{aligned}
  {\rm FF}_{n} = \frac{\sum^{N_{\rm gen}}_{k=1} \left|c_{n}\mathcal{A}^k_{n}\right|^{2}}{\sum^{N_{\rm gen}}_{k=1} \left|\mathcal{M}^k\right|^{2}}\,, \label{Fit-Fraction-Definition}
\end{aligned}\end{eqnarray}
where $N_{\rm gen}$ is the number of phase space MC signal events at generator level.
The sum of these FFs is generally not unity due to net constructive or destructive
interference. Interference IN between the $n^{\rm{th}}$ and $n^{\prime\rm{th}}$
amplitudes is defined as
\begin{eqnarray}\begin{aligned}
    {\rm IN}_{nn^{\prime}} = \frac{\sum^{N_{\rm gen}}_{k=1} 2{\rm Re}[c_{n}c^{*}_{n^{\prime}}\mathcal{A}^k_{n}\mathcal{A}^{k*}_{n^{\prime}}]}{\sum^{N_{\rm gen}}_{k=1} \left|\mathcal{M}^k\right|^{2}}\,. \label{interferenceFF-Definition}
\end{aligned}\end{eqnarray}
In order to determine the statistical uncertainties of FFs, the amplitude
coefficients are randomly selected by a Gaussian-distributed set by the fit results
according to their uncertainties and the covariance matrix.  The distribution of each FF is fitted with
a Gaussian function whose width is then taken as the uncertainty of this FF.

\begin{figure*}[!htbp]
 \centering
    \includegraphics[width=0.45\textwidth]{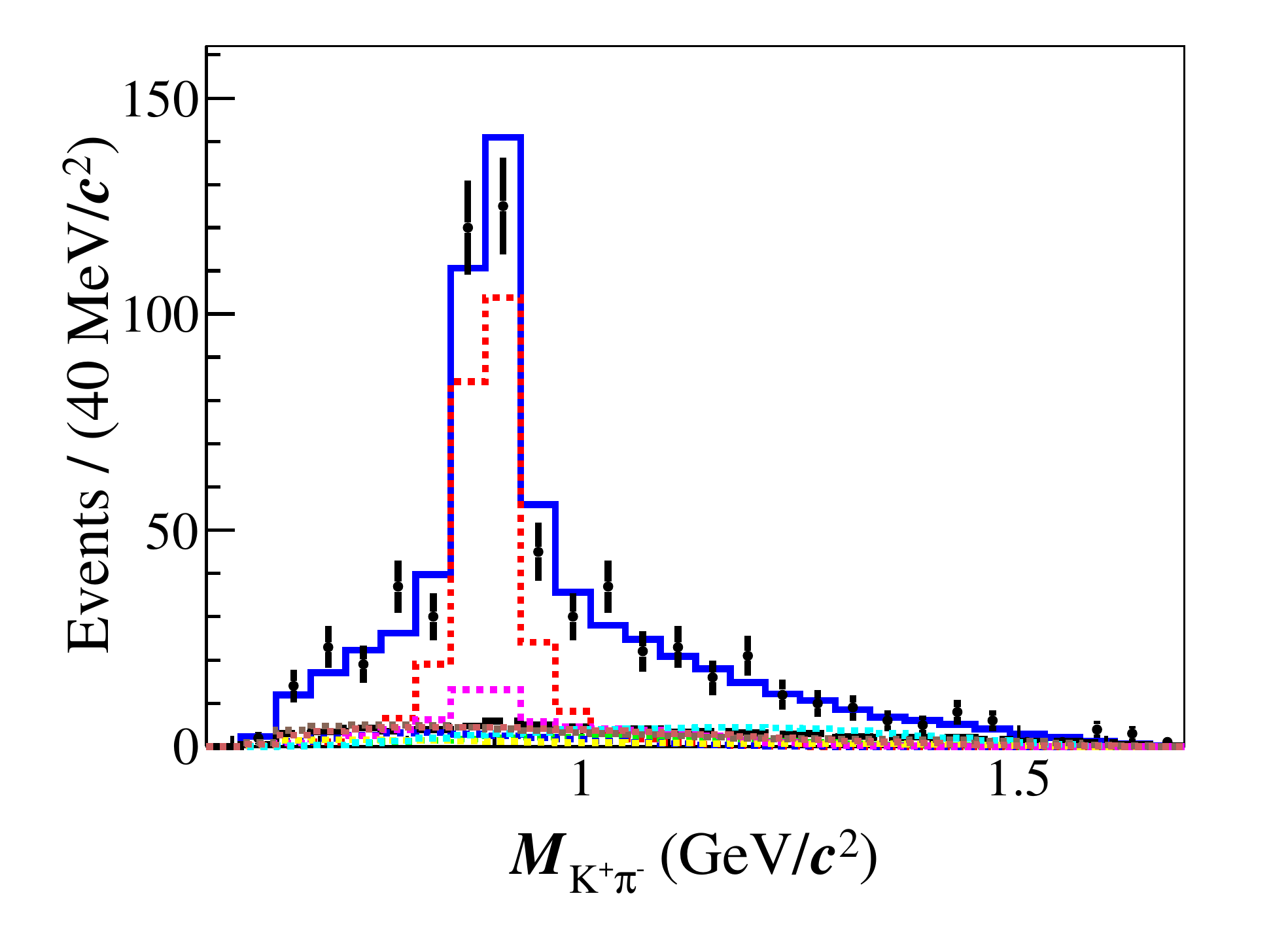}
    \includegraphics[width=0.45\textwidth]{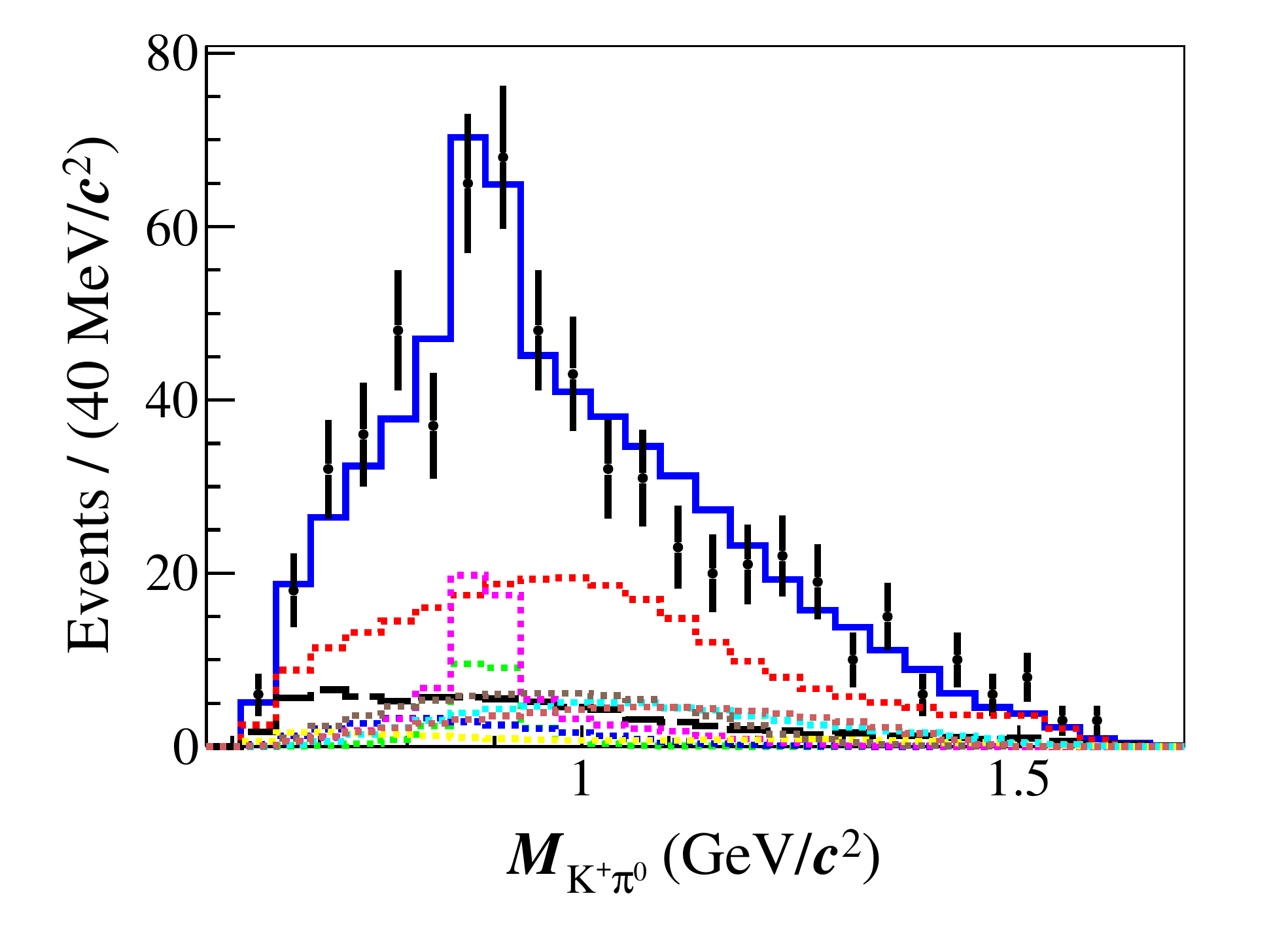}
    \includegraphics[width=0.45\textwidth]{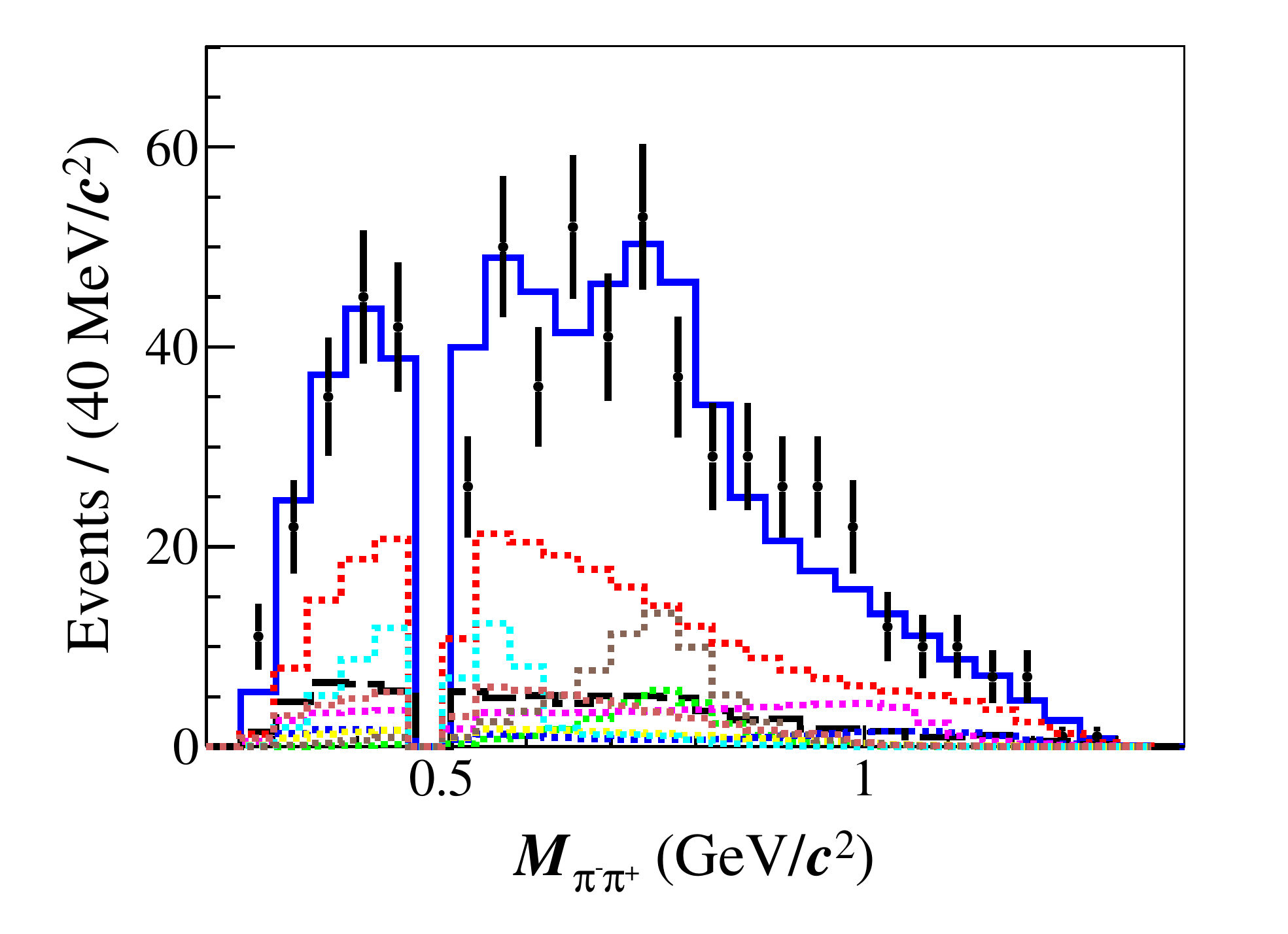}
    \includegraphics[width=0.45\textwidth]{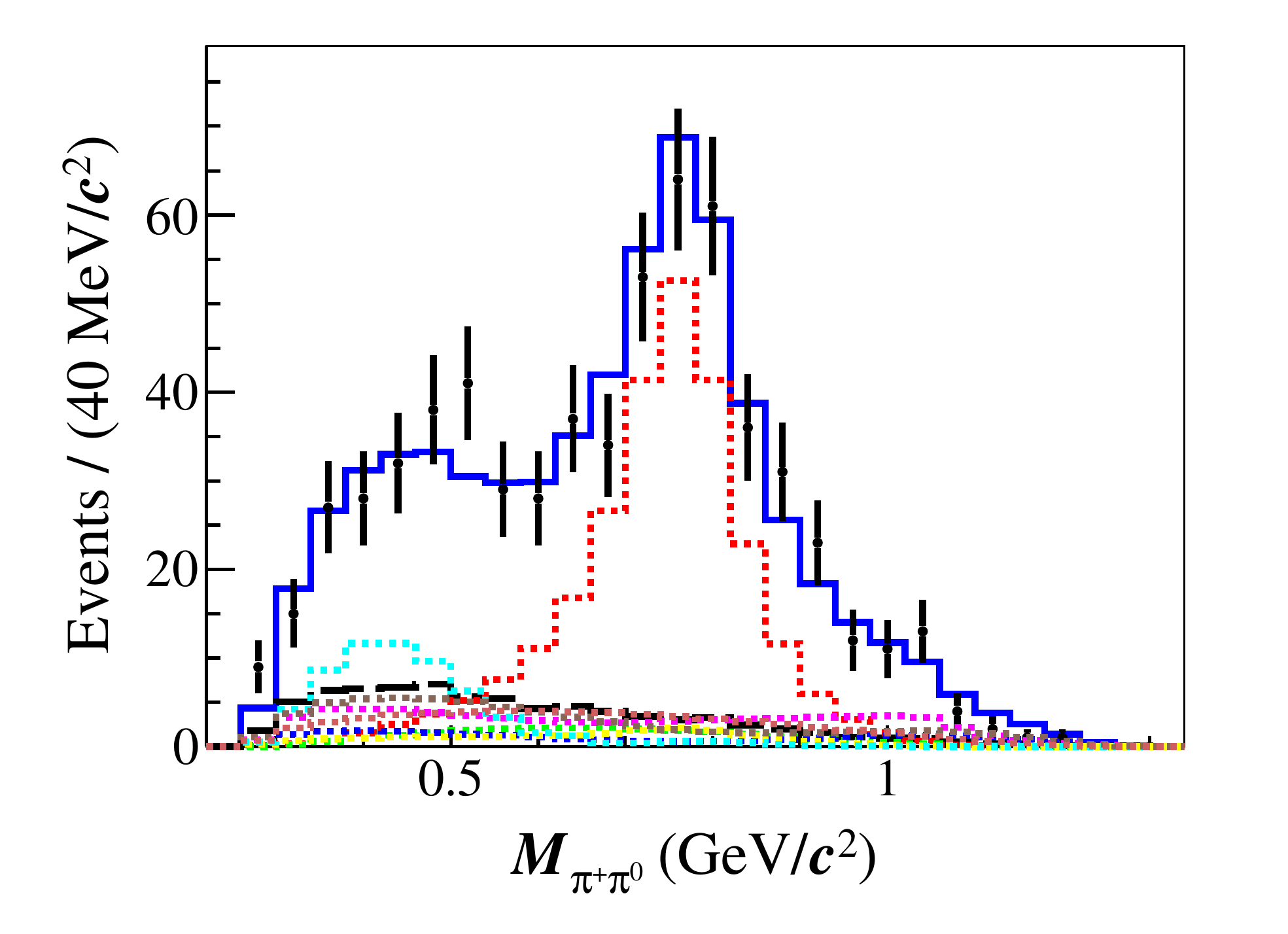}
    \includegraphics[width=0.45\textwidth]{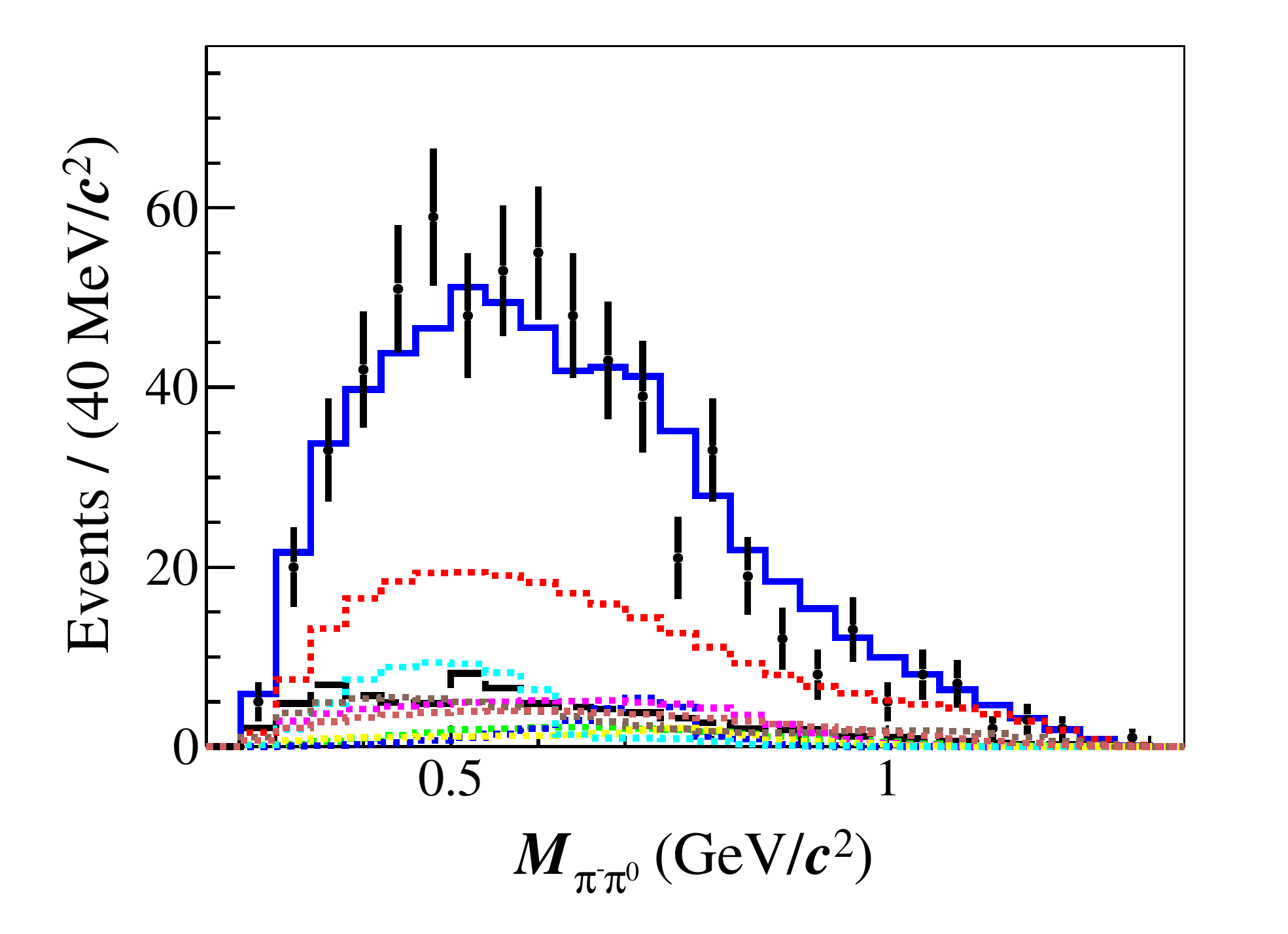}
    \includegraphics[width=0.45\textwidth]{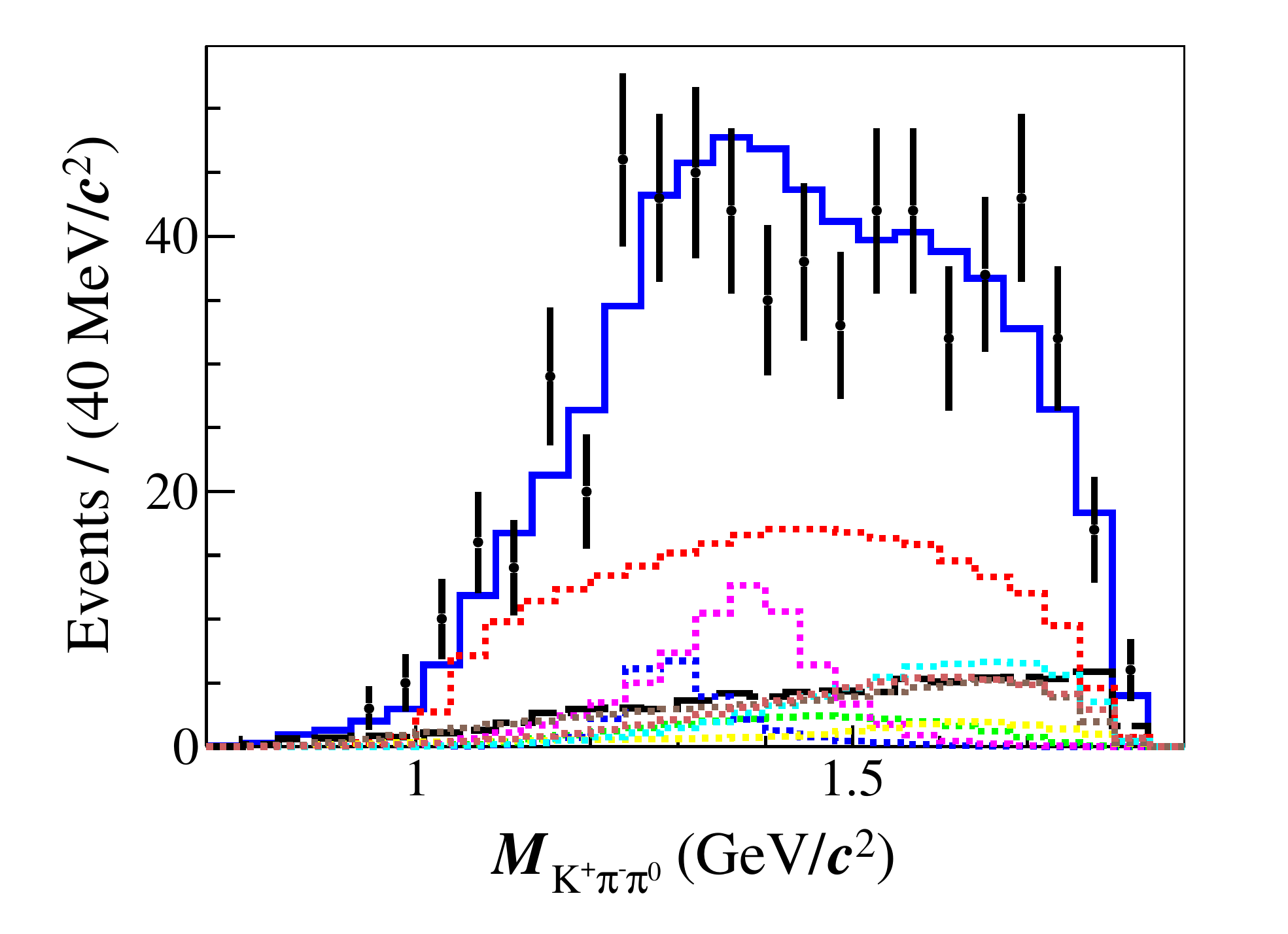}
    \includegraphics[width=0.45\textwidth]{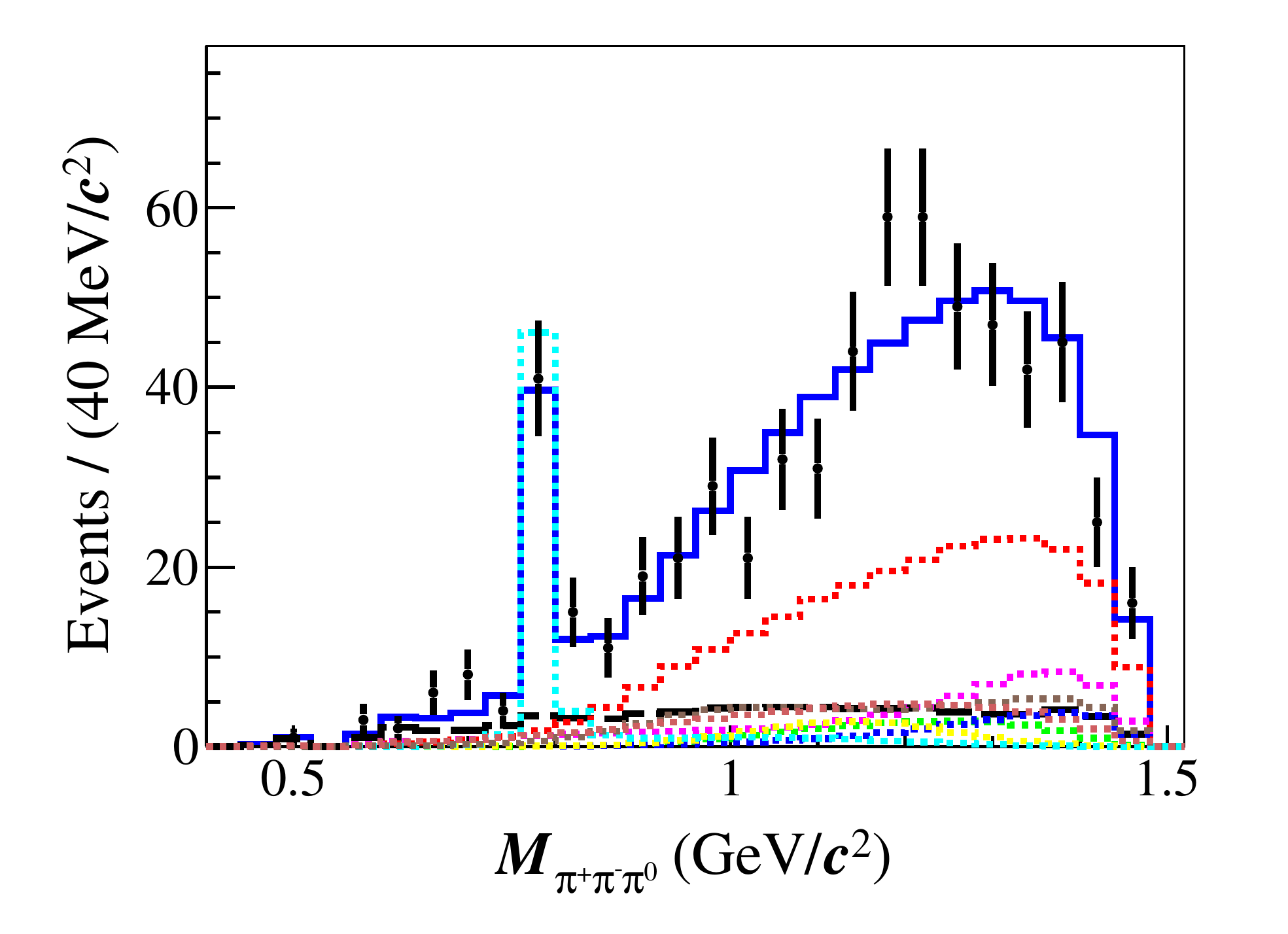}
    \includegraphics[width=0.45\textwidth]{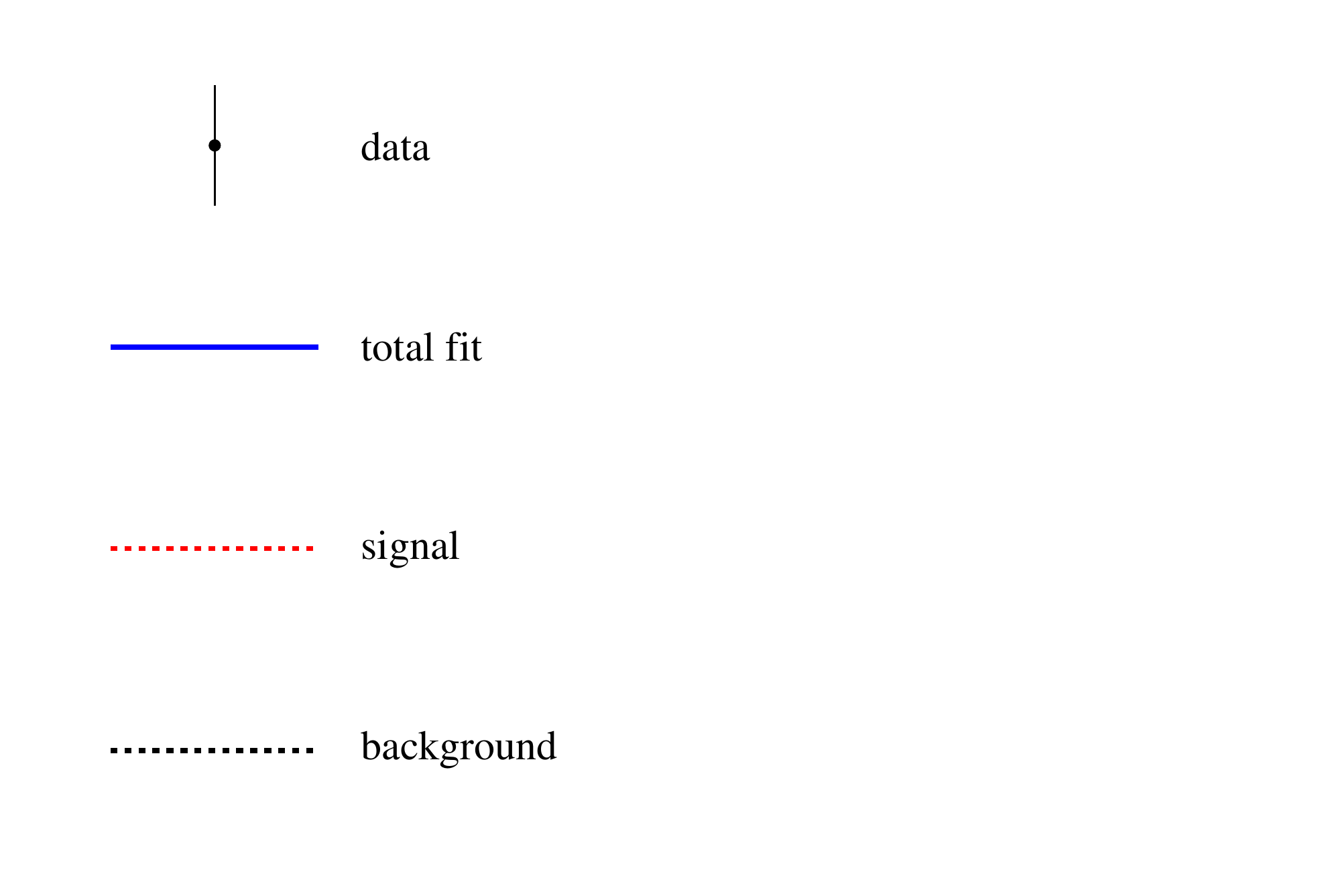}
 \caption{Selected mass projections of the nominal fit. The data samples at $\sqrt{s}= 4.178-4.226$~GeV are
   represented by points with error bars, the fit results by the solid blue
   lines, and the background estimated from the inclusive MC sample by the black
   dashed lines. Coloured curves show different components of the nominal fit. The total fit is not necessarily equal to the sum of the components due to interference effects.
}
 \label{dalitz-projection}
\end{figure*}

%------------------------------------------------------------------------------
The phases, FFs, and statistical significances(Stat.Signi) for different amplitudes are listed in
Table~\ref{fit-result}. The mass projections of the nominal fit are shown in Fig.~\ref{dalitz-projection}.

\begin{table*}[htbp]\small
\setlength{\abovecaptionskip}{0.cm}
\setlength{\belowcaptionskip}{-0.1cm}
    \begin{center}
      \begin{tabular}{|l r@{ $\pm$ }c@{ $\pm$ }c r@{ $\pm$ }c@{ $\pm$ }c c|}
      \hline
      Amplitude                           &\multicolumn{3}{c}{Phase~(rad)}  &\multicolumn{3}{c}{FF~(\%)} &Stat.Signi~($\sigma$)\\
      \hline
         $D_s^{+}[S]\to K^{*0}\rho^+$        &\multicolumn{3}{c}{0.0~(fixed)}   & 14.5 &2.2 &0.6            &>10\\
         $D_s^{+}[P]\to K^{*0}\rho^+$        &2.09 & 0.12 & 0.03          & 26.0 &2.5 &1.1               &>10 \\
         $D_s^{+}\to K^{*0}\rho^+$           &\multicolumn{3}{c}{-}        & 40.5 & 2.8 & 1.5              &>10 \\
        \hline
         $D_s^{+}[P]\to K^{*+}\rho^0$        &2.42 & 0.21 & 0.04           & 4.3  & 1.1 & 0.6              &6.8 \\
        \hline
         $D_s^{+}\to K^+\omega$              &0.57 & 0.23 & 0.10           & 9.7 & 1.5 & 0.6               &>10 \\
        \hline
         $D_s^{+}\to K_1(1270)^0(K^+ \rho^-)[S]\pi^+$    &1.80 & 0.24 & 0.08   & 4.0  & 1.2 & 0.6          &5.5\\
        \hline
         $D_s^{+}\to K_1(1400)^0(K^{*+}\pi^-)[S]\pi^+$   &$-1.61$ & 0.17 & 0.05  & 5.6  & 0.9 & 0.2     &-\\
         $D_s^{+}\to K_1(1400)^0(K^{*0}\pi^0)[S]\pi^+$   &$-1.61$ & 0.17 & 0.05  & 6.1  & 0.9 & 0.2     &-\\
         $D_s^{+}\to K_1(1400)^0(K^{*}\pi)[S]\pi^+$      &\multicolumn{3}{c}{-}                       & 11.3 & 1.8 & 0.4     &8.9\\
        \hline
         $D_s^{+}\to K^+a_1(1260)^0(\rho^+\pi^-)[S]$         &$-1.19$ & 0.25 & 0.22   & 1.9 & 0.7  & 0.9     &-\\
         $D_s^{+}\to K^+a_1(1260)^0(\rho^-\pi^+)[S]$         &$-1.19$ & 0.25 & 0.22   & 1.9 & 0.7  & 0.9     &-\\
         $D_s^{+}\to K^+a_1(1260)^0(\rho\pi)[S]$             &\multicolumn{3}{c}{-}                    & 3.3 & 1.2  & 1.5     &3.8\\
        \hline
         $D_s^{+}[S]\to (K^+\pi^0)_V\rho^0$                  &1.02 & 0.16 & 0.08  & 10.4 & 2.0 & 0.6     &6.6\\
        \hline
         $D_s^{+}\to (K^+\pi^0)_{\rm S-wave}(\pi^+\pi^-)_{\rm S-wave}$                &$-2.87$ & 0.17 & 0.06   & 9.5  & 2.2 & 0.9     &6.0\\
      \hline
    \end{tabular}
    \end{center}
  \caption{Phases, FFs, and statistical significances for different amplitudes.
      Groups of related amplitudes are separated by horizontal lines. The last row
      of each group gives the total fit fraction of the above components with interferences considered.
      The amplitudes $K_1(1400)^0$ and $a_1(1260)^0$ are constructed by two sub-amplitudes with fixed relations (see Appendix~\ref{tested_amplitude}).
      The $K^{*0(+)}$ resonance decays to $K^+\pi^{-(0)}$. The $\rho^{+(0)}$ resonance decays to $\pi^+\pi^{0(-)}$. The $\omega$ resonance decays to $\pi^+\pi^-\pi^0$. Note that $K^*$ indicates $K^{*+}$ and $K^{*0}$, $\rho$ indicates $\rho^{+}$ and $\rho^{0}$.
      The first and second uncertainties in the phases and FFs are statistical and systematic,
      respectively. The total FF is 93.9$\%$.}
    \label{fit-result}
\end{table*}

%------------------------------------------------------------------------------
\subsection{Systematic uncertainties for the amplitude analysis}
\label{sec:PWA-Sys}
The systematic uncertainties for the amplitude analysis, summarised in Table~\ref{pwa:sys},
are now detailed.

\begin{itemize}
\item[\lowercase\expandafter{\romannumeral1}]
Amplitude model: \\
The masses and widths of resonances are varied by their corresponding uncertainties~\cite{PDG,K1270}. The GS lineshape of $\rho$ is replaced with the RBW formula. The coupling constants of the $\pi\pi$ S-wave model are varied within their uncertainties given in Ref.~\cite{PPs4}. The changes of the phases and FFs are assigned as the associated systematic uncertainties.
%The uncertainties of lineshape of $K\pi$ S-wave model are estimated by replacing the lineshape from BABAR to the K-matrix formula~\cite{Kx}.
Since replacing the lineshape of the $K\pi$ S-wave model from BABAR with the $K$-matrix formula~\cite{Kx} results in different normalisation factors, the effect on the phase of the amplitude related to $K\pi$ S-wave is not considered for this source.
%The coupling constants of $K\pi$ S-wave model are varied within their uncertainties given by Ref.~\cite{KPs2}. In the baseline fit the $s_0^{prod}$ parameter appearing in the Eq.~\ref{pipis} is varied between [-1, -5] GeV$^2/c^4$~\cite{PPs4}.
\item[\lowercase\expandafter{\romannumeral2}]
$R$ values: \\
We assume the distribution of values for barrier effective radius~($R$), as defined in Sec.~\ref{sec:barrier}, as a uniform distribution. The systematic uncertainties associated with $R$ are estimated by repeating the fit procedure by varying the $R$ of both the intermediate state and $D_s^+$ mesons by $R$/$\sqrt{12} \approx 1$~GeV$^{-1}$.
\item[\lowercase\expandafter{\romannumeral3}]
Background: \\
        The uncertainty from background size is studied by varying the fractions of signal (equivalent to the fractions of background), i.e. $w^i$ in Eq.~\ref{likelihood3}, within their corresponding statistical uncertainties. Another source is the simulation of background shapes. First, alternative MC shapes where the relative fractions of the dominant backgrounds from $e^+e^- \to q\bar{q}$ ($q=u,d,s$) and non-$D_s^{*\pm}D_s^{\mp}$ open charm processes are varied by the statistical uncertainties of their cross sections are used.
Second, the background PDF is extracted using other five variable input combinations with varied smoothing parameters of RooNDKeysPDF~\cite{Verkerke}.
%TODO: I cannot work on this sentence without enough knowledge on this analysis, e.g., why there are *five* combinations? Please explain here.
\item[\lowercase\expandafter{\romannumeral4}]
Simulation effects: \\
To estimate the uncertainties caused by $\gamma_{\epsilon}$, as defined in Eq.~\ref{pwa:gamma}, an amplitude fit is performed by varying efficiencies of PID, tracking and $\pi^0$ reconstruction according to their uncertainties.
\item[\lowercase\expandafter{\romannumeral5}]
Fit bias:\\
The uncertainty from the fit process is evaluated by studying 600 signal MC samples with the size equal to the data sample size that are generated to check the pull. The pull variables, $(V_{\rm input}-V_{\rm fit})/\sigma_{\rm fit}$,  are defined to evaluate the corresponding uncertainty, where $V_{\rm input}$ is the input value in the generator, $V_{\rm fit}$ and $\sigma_{\rm fit}$ are the output value and the corresponding statistical uncertainty, respectively. Expected to be
the standard normal distribution for an unbiased fit, the distributions of
pull values for the 600 sets of sample are fitted with a Gaussian function.
        The fitted mean values for the pulls of FFs of $D_s^+[P] \to K^{*0}\rho^+$ and $D_s^+ \to (K^+\pi^0)_{\rm S-wave}(\pi^+\pi^-)_{\rm S-wave}$ deviate from zero by larger than 3 times of the standard deviation.  We correct all resonances' FFs and phases by the fitted mean values, and assign the uncertainty of the fitted mean values as the corresponding systematic uncertainties.
\end{itemize}

\begin{table}[t]\small
\setlength{\abovecaptionskip}{0.3cm}
\setlength{\belowcaptionskip}{0.cm}
	\centering
	\begin{tabular}{|lccccccc|}
	\hline
		\   &\multicolumn{7}{c|}{Source}    \\
	\hline
		Amplitude                                             &\    &\lowercase\expandafter{\romannumeral1}    &\lowercase\expandafter{\romannumeral2}    &\lowercase\expandafter{\romannumeral3} &\lowercase\expandafter{\romannumeral4}   &\lowercase\expandafter{\romannumeral5}  &Total\\
	\hline
		$D_s^{+}[S]\to K^{*0}\rho^+$                      & FF      &0.09 &0.10 &0.19 &0.01 &0.06 &0.25\\
\multirow{2}{*}{$D_s^{+}[P]\to K^{*0}\rho^+$}           & $\phi$  &0.20 &0.02 &0.07 &0.01 &0.06 &0.24\\
 	                                                         & FF      &0.10 &0.08 &0.40 &0.03 &0.06 &0.44\\
		$D_s^{+}\to K^{*0}\rho^+$                         & FF      &0.14 &0.02 &0.50 &0.02 &0.06 &0.53\\
		\hline
		\multirow{2}{*}{$D_s^{+}[P]\to K^{*+}\rho^0$}     & $\phi$  &0.08 &0.04 &0.11 &0.01 &0.05 &0.16\\
 			                                                   & FF      &0.15 &0.32 &0.39 &0.01 &0.06 &0.54\\
		\hline
		\multirow{2}{*}{$D_s^{+}\to K^+\omega$}               & $\phi$  &0.38 &0.11 &0.13 &0.01 &0.05 &0.42\\
							                                       &FF        &0.09 &0.19 &0.31 &0.01 &0.06 &0.38\\
		\hline
		\multirow{2}{*}{$D_s^{+}\to K_1(1270)^0(K^+ \rho^-)[S]\pi^+$}  & $\phi$ &0.23 &0.11 &0.17 &0.01 &0.06 &0.32\\
									                                        & FF     &0.13 &0.05 &0.41 &0.01 &0.06 &0.46\\
		\hline
		\multirow{2}{*}{$D_s^{+}\to K_1(1400)^0(K^{*+}\pi^-)[S]\pi^+$}  & $\phi$   &0.24 &0.08 &0.21 &0.01 &0.06 &0.34\\
														                              & FF       &0.18 &0.08 &0.13 &0.01 &0.06 &0.24\\

		\multirow{2}{*}{$D_s^{+}\to K_1(1400)^0(K^{*0}\pi^0)[S]\pi^+$}  & $\phi$   &0.24 &0.08 &0.21 &0.01 &0.06 &0.34\\
														                              & FF       &0.20 &0.06 &0.13 &0.02 &0.06 &0.25\\
		$D_s^{+}\to K_1(1400)^0(K^{*}\pi)[S]\pi^+$                                    & FF       &0.16 &0.09 &0.12 &0.01 &0.06 &0.23\\
		\hline
		\multirow{2}{*}{$D_s^{+}\to K^+a_1(1260)^0(\rho^+\pi^-)[S]$}     & $\phi$   &0.83 &0.16 &0.19 &0.01 &0.06 &0.87\\
													                                 & FF       &0.79 &0.19 &1.01 &0.01 &0.06 &1.31\\
		\multirow{2}{*}{$D_s^{+}\to K^+a_1(1260)^0(\rho^-\pi^+)[S]$}     & $\phi$   &0.83 &0.16 &0.19 &0.01 &0.06 &0.87\\
													                                 & FF       &0.79 &0.19 &1.01 &0.01 &0.06 &1.31\\
		$D_s^{+}\to K^+a_1(1260)^0(\rho\pi)[S]$                                      & FF       &0.77 &0.16 &1.00 &0.02 &0.06 &1.28\\
		\hline
		\multirow{2}{*}{$D_s^{+}[S]\to (K^+\pi^0)_{V}\rho^0$}              & $\phi$   &0.08 &0.02 &0.51 &0.02 &0.06 &0.53\\
								                                               & FF      &0.15 &0.19 &0.13 &0.05 &0.06 &0.30\\
		\hline
		\multirow{2}{*}{$D_s^{+}\to (K^+\pi^0)_{\rm S-wave}(\pi^+\pi^-)_{\rm S-wave}$}   & $\phi$   &0.24 &0.21 &0.04 &0.01 &0.04 &0.33\\
										                                         &  FF       &0.14 &0.05 &0.35 &0.01 &0.07 &0.40\\
	\hline
	\end{tabular}
	\caption{Systematic uncertainties on the phases and FFs for different amplitudes in units of the corresponding statistical uncertainties.
	(\lowercase\expandafter{\romannumeral1}) Amplitude model, (\lowercase\expandafter{\romannumeral2}) Effective radius, (\lowercase\expandafter{\romannumeral3}) Background, (\lowercase\expandafter{\romannumeral4}) Experimental effects, (\lowercase\expandafter{\romannumeral5}) Fit bias.}
	\label{pwa:sys}
\end{table}
%------------------------------------------------------------------------------

\section{Branching fraction measurement}
\label{BFSelection}
On top of the selection criteria described in Sec.~\ref{ST-selection}, the momenta of all pions are
further required to be greater than 100 MeV/$c$ to exclude soft pions from $D^{*}$ decays. The best tag
candidate is chosen with $M_{\rm rec}$ closest to $m_{D_{s}}$ %the known $D_{s}^{\pm}$ mass~\cite{PDG}
if there are multiple ST candidates. The yields for
various tag modes are obtained from the fits to the corresponding $M_{\rm tag}$
distributions and the results are summarised in Table~\ref{ST-eff}. As an example, the fits to the
data sample at $\sqrt s=4.178$~GeV are shown in Fig.~\ref{fit:Mass-data-Ds_4180}.
In the fits, the signal is modeled by an MC-simulated shape convolved with a
Gaussian function to take into account the data-MC resolution difference. The
background is described by a second-order Chebychev polynomial.
Inclusive MC studies show that there is no significant peaking background in any tag mode,
except for $D^{-} \to K_{S}^{0} \pi^-$ and $D_{s}^{-} \to \eta\pi^+\pi^-\pi^-$ faking
the $D_{s}^{-} \to K_{S}^{0} K^-$ and $D_{s}^{-} \to \pi^-\eta^{\prime}$ tags,
respectively. Therefore, the MC-simulated shapes of these two
peaking background sources, with the yields included as free parameters, are added to the fits, respectively.
\begin{table*}[t]
\setlength{\abovecaptionskip}{0.cm}
\setlength{\belowcaptionskip}{0.cm}
    \begin{center}
      \begin{tabular}{|lccc|}
        \hline
        Tag mode                                    & $N_{\rm ST}$(I)           & $N_{\rm ST}$(II)       & $N_{\rm ST}$(III)      \\
        \hline
        $D_{s}^{-}\to K_{S}^{0}K^{-}$               & $\phantom{0}31941\pm312$   & $18559\pm261$            & $\phantom{0}6582\pm160$ \\
        $D_{s}^{-}\to K^{+}K^{-}\pi^{-}$            & $137240\pm614$             & $81286\pm505$            & $28439\pm327$           \\
        $D_{s}^{-}\to K^{+}K^{-}\pi^{-}\pi^{0}$     & $\phantom{0}39306\pm799$   & $23311\pm659$            & $\phantom{0}7785\pm453$           \\
        $D_{s}^{-}\to K_{S}^{0}K^{+}\pi^{-}\pi^{-}$ & $\phantom{0}15719\pm289$   & $\phantom{0}8948\pm231$  & $\phantom{0}3263\pm172$ \\
        $D_{s}^{-}\to \pi^{-}\eta_{\gamma\gamma}$   & $\phantom{0}17940\pm402$   & $10025\pm339$            & $\phantom{0}3725\pm252$ \\
        $D_{s}^{-}\to \pi^{-}\eta^{\prime}_{\pi^+\pi^-\eta_{\gamma\gamma}}$         & $\phantom{00}7759\pm141$   & $\phantom{0}4428\pm111$  & $\phantom{0}1648\pm74\phantom{0}$ \\
        \hline
      \end{tabular}
    \end{center}
  \caption{The ST yields for the data samples collected at $\sqrt{s} =$ (I) 4.178~GeV, (II) 4.189-4.219~GeV,
    and (III) 4.226~GeV. The uncertainties are statistical.}\label{ST-eff}
\end{table*}

\begin{figure*}[t]
\setlength{\abovecaptionskip}{0.cm}
\setlength{\belowcaptionskip}{0.cm}
\begin{center}
	\includegraphics[width=0.90\textwidth]{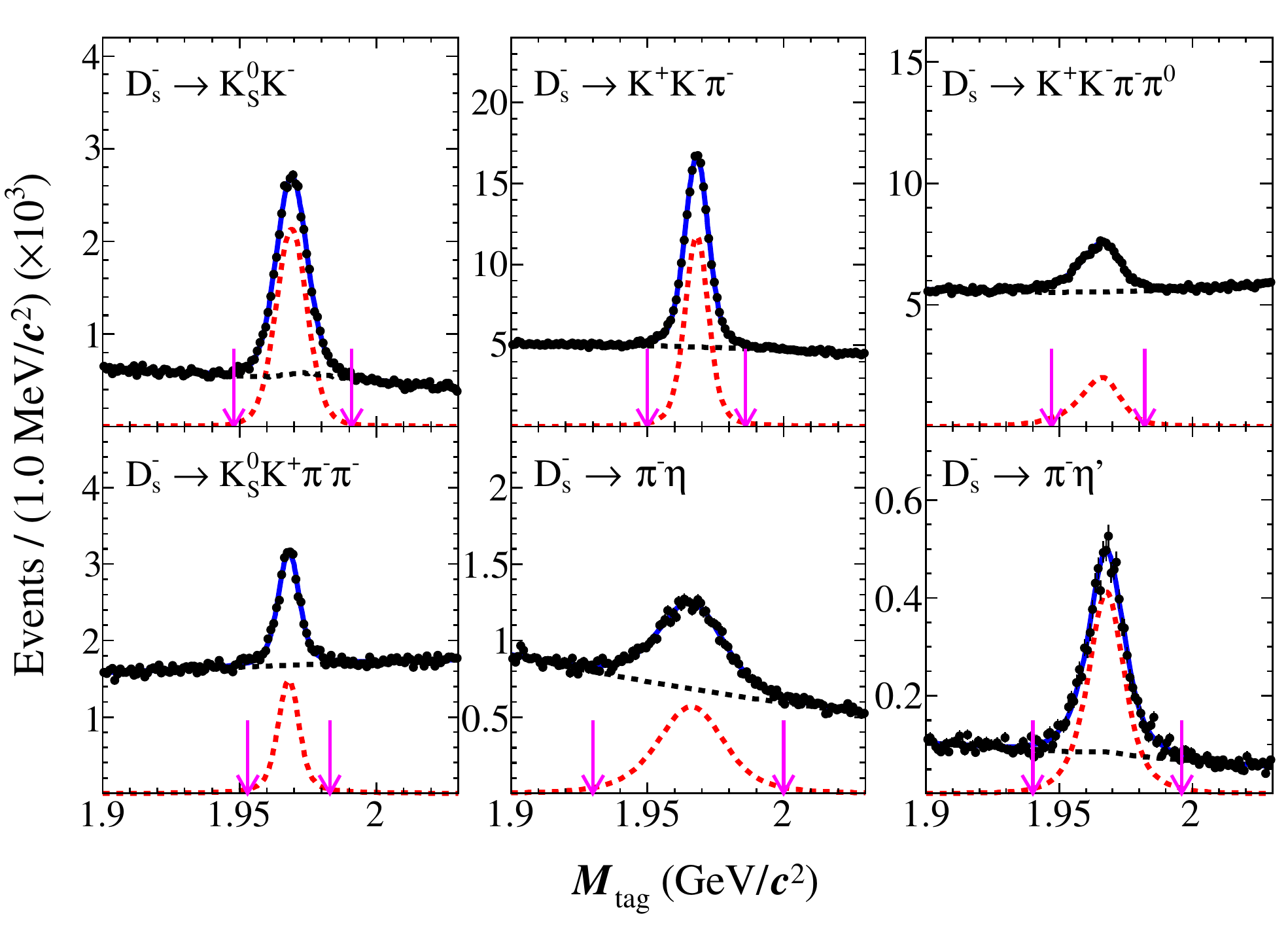}
\caption{Fits to the $M_{\rm tag}$ distributions of the ST candidates
         from the data sample at $\sqrt{s}=4.178$~GeV. The points with
         error bars are data, the blue solid lines are the total fits, and the black
         dashed lines are the fitted background. The pairs of red arrows denote the
         signal regions.
         }
\label{fit:Mass-data-Ds_4180}
\end{center}
\end{figure*}

Once a tag mode is identified, we select the signal decay
$D_{s}^{+} \to K^+\pi^{+}\pi^{-}\pi^{0}$. In the case of multiple candidates, the
DT candidate with the average mass, $(M_{\rm sig}+M_{\rm tag})/2$,
closest to $m_{D_{s}}$ %the known $D_{s}^{\pm}$ mass~\cite{PDG}
is retained.

For a given single tag mode, the ST and DT yields can be written as:
\begin{eqnarray}\begin{aligned}
  N_{\text{tag}}^{\text{ST}} = 2N_{D_{s}^{+}D_{s}^{-}}\mathcal{B}_{\text{tag}}\epsilon_{\text{tag}}^{\text{ST}}\,, \label{eq-ST}
\end{aligned}\end{eqnarray}
\begin{equation}
  N_{\text{tag,sig}}^{\text{DT}}=2N_{D_{s}^{+}D_{s}^{-}}\mathcal{B}_{\text{tag}}\mathcal{B}_{\text{sig}}\epsilon_{\text{tag,sig}}^{\text{DT}}\,,
  \label{eq-DT}
\end{equation}
where $N_{D_{s}^{+}D_{s}^{-}}$ is the total number of $D_{s}^{*\pm}D_{s}^{\mp}$
pairs produced in the data samples, $N_{\text{tag}}^{\text{ST}}$ is
the ST yield for the tag mode; $N_{\text{tag,sig}}^{\text{DT}}$ is the DT yield;
$\mathcal{B}_{\text{tag}}$ and $\mathcal{B}_{\text{sig}}$ are the BFs of the
tag and signal modes, respectively; $\epsilon_{\text{tag}}^{\text{ST}}$ is the
ST efficiency to reconstruct the tag mode; and $\epsilon_{\text{tag,sig}}^{\text{DT}}$
is the DT efficiency to reconstruct both the tag and signal modes. Summing over tag modes and sample groups gives the total DT yield:
\begin{eqnarray}
\begin{aligned}
  \begin{array}{lr}
    N_{\text{total}}^{\text{DT}}=\Sigma_{\alpha, i}N_{\alpha,\text{sig},i}^{\text{DT}}   = \mathcal{B}_{\text{sig}}
 \Sigma_{\alpha, i}2N^i_{D_{s}^{+}D_{s}^{-}}\mathcal{B}_{\alpha}\epsilon_{\alpha,\text{sig}, i}^{\text{DT}}\,,
  \end{array}
  \label{eq-DTtotal}
\end{aligned}
\end{eqnarray}
where $\alpha$ represents tag modes in the $i^{\rm th}$ sample group.
Therefore, the BF of the signal decay can be determined by
\begin{eqnarray}\begin{aligned}
  \mathcal{B}_{\text{sig}} =
  \frac{N_{\text{total}}^{\text{DT}}}{\mathcal{B}_{\pi^0\to\gamma\gamma}\begin{matrix}\sum_{\alpha, i} N_{\alpha, i}^{\text{ST}}\epsilon^{\text{DT}}_{\alpha,\text{sig},i}/\epsilon_{\alpha,i}^{\text{ST}}\end{matrix}}\,,\label{eq:bf}
\end{aligned}\end{eqnarray}
where $N_{\alpha,i}^{\text{ST}}$ and $\epsilon_{\alpha,i}^{\text{ST}}$ are
obtained from the data and inclusive MC samples, respectively, while
$\epsilon_{\alpha,\text{sig},i}^{\text{DT}}$ is determined with signal MC
samples generated based on our amplitude analysis. The branching ratio
$\mathcal{B}_{\pi^0\to\gamma\gamma}$ has been introduced as it is not included in the MC generation.

\begin{figure}[!htbp]
\setlength{\abovecaptionskip}{0.cm}
\setlength{\belowcaptionskip}{0.cm}
  \centering
  \includegraphics[width=0.6\textwidth]{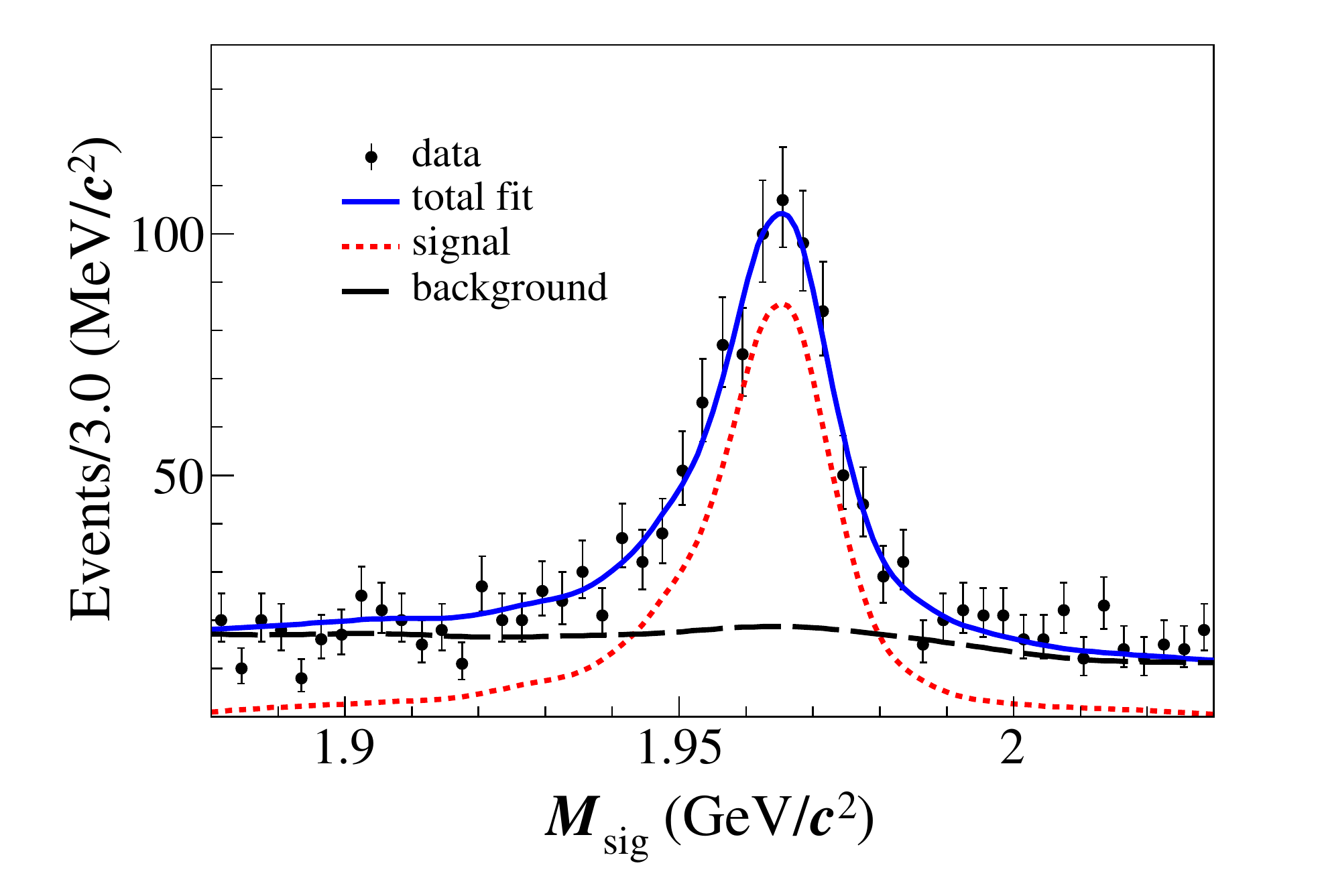}
 \caption{Fit to the $M_{\rm sig}$ distribution of the DT candidates from the
   data samples  at $\sqrt{s}= 4.178$-$4.226$~GeV. The data are represented by
   points with error bars, the total fit by the blue solid line, and the fitted
   signal and background by the red dotted and black dashed
   lines, respectively.
 }
  \label{DT-fit}
\end{figure}

The DT yield $N^{\rm DT}_{\rm total}$ is found to be $776\pm43$ from the fit to the $M_{\rm sig}$ distribution of the selected $D_{s}^{+} \to K^+\pi^{+}\pi^{-}\pi^{0}$ candidates. The fit result is shown in Fig.~\ref{DT-fit}, where the signal shape is modeled by an MC-simulated shape convolved with a Gaussian function to take into account the data-MC resolution difference. The background shape is derived from the inclusive MC sample.
After correcting for the differences in $K^+$ and $\pi^{\pm}$ tracking, PID and $\pi^{0}$ reconstruction efficiencies between data and
MC simulation, we determine the BF of $D_{s}^{+} \to K^+\pi^{+}\pi^{-}\pi^{0}$
to be $(9.75\pm0.54_{\rm stat.}\pm0.17_{\rm syst.})\times 10^{-3}$ according to Eq.~\ref{eq:bf}.

The BFs for the charge-conjugated modes $D_{s}^{+} \to K^+\pi^{+}\pi^{-}\pi^{0}$ and
$D_{s}^{-} \to K^-\pi^{-}\pi^{+}\pi^{0}$, which are labeled as ${\mathcal B}(D^+_s)$ and ${\mathcal B}(D^-_s)$, are measured to be ($9.10\pm0.71_{\rm{stat.}}\pm0.16_{\rm{syst.}}) \times 10^{-3}$ and ($10.39\pm0.79_{\rm{stat.}}\pm0.18_{\rm{syst.}}) \times 10^{-3}$, respectively.
The asymmetry of the BFs, $A_{CP} = \frac{\mathcal{B}(D_s^+)-\mathcal{B}(D_s^-)}{\mathcal{B}(D_s^+)+\mathcal{B}(D_s^-)}$, is determined to be $(6.5\pm5.4_{\rm{stat.}}\pm0.7_{\rm{syst.}})\%$. No significant $CP$ violation is observed with the current sample size. Note that the systematic uncertainties due to pion tracking and PID, $\pi^0$ reconstruction are canceled in the $A_{CP}$ calculation.

%------------------------------------------------------------------------------
%------------------------------------------------------------------------------
The systematic uncertainties in the BF measurement are discussed as follow.
\begin{itemize}
%\item Signal shape.
%The systematic uncertainty due to the signal shape is studied by repeating the
%fit without the convolved Gaussian.
  \item Background shape:\\
Since the $e^+e^-\to q\bar q$ and non-$D_{s}^{*\pm}D_{s}^{\mp}$
open charm processes are the major background sources, an alternative MC-simulated background shape
is obtained by varying the relative fractions of the background components from these processes by the statistical uncertainties of their cross sections.
%This $30\%$ is the statistical uncertainties of the cross section of $q\bar{q}$
%and non-$D_{s}^{*+}D_{s}^{-}$ open charm in the data sample.
The largest change, 0.5\%, is taken as the related uncertainty.
\item Tracking and PID efficiency:\\
The $\pi^{\pm}$ and $K^{+}$ tracking and PID efficiencies are studied with the control samples of
        $e^+e^-\to K^+K^-K^+K^-$, $K^+K^-\pi^+\pi^-(\pi^0)$, and $\pi^+\pi^-\pi^+\pi^-(\pi^0)$ decays. The data-MC tracking and PID efficiency ratios of $\pi^{+}(\pi^{-})$ are 1.001 $\pm$ 0.003 (0.997 $\pm$ 0.003) and 0.998 $\pm$ 0.002 (0.998 $\pm$ 0.002), respectively. The data-MC tracking and PID efficiency ratios of $K^{+}(K^{-})$ are 1.004 $\pm$ 0.006 (1.005 $\pm$ 0.006) and 0.998 $\pm$ 0.002 (0.998 $\pm$ 0.002), respectively. After correcting the MC efficiencies to data by these factors, the statistical uncertainties of the correction parameters are assigned to the systematic uncertainties associated with tracking and PID efficiencies. They are determined to be 0.3\% (0.2\%) for each $\pi^{+}$ ($\pi^{-}$), and  0.6\% (0.2\%) for each $K^{+}$ ($K^{-}$), respectively.
\item $\pi^{0}$ reconstruction:\\
The $\pi^{0}$ reconstruction efficiency is
investigated by using a control sample of the process
$e^+e^-\to K^+K^-\pi^+\pi^-\pi^0$.
The data-MC efficiency ratio for $\pi^0$ reconstruction is estimated to be
$0.995 \pm 0.008$. After correcting the efficiency by this factor,
we assign 0.8\% as the systematic uncertainty.
%The efficiency difference of data and MC samples is then determined to be
%2.0\% per $\pi^0$.
\item MC sample size:\\
The uncertainty due to the limited MC sample size is obtained by
$\sqrt{\begin{matrix} \sum_{\alpha} (f_{\alpha}\frac{\delta_{\epsilon_{\alpha}}}{\epsilon_{\alpha}}\end{matrix}})^2$,
where $f_{\alpha}$ is the tag yield fraction, and $\epsilon_{\alpha}$ and
$\delta_{\epsilon_{\alpha}}$ are the signal efficiency and the corresponding
uncertainty of tag mode $\alpha$, respectively. The uncertainty corresponding to MC statistics is 0.2\%.

%------------------------------------------------------------------------------
\item Amplitude analysis model:\\
The uncertainty arising from the amplitude analysis model is estimated by varying the model parameters based on their error matrix.
The distribution of 600 efficiencies resulting from this variation is fitted by a
Gaussian function. The fitted width divided by the mean value, 0.4\%, is taken as an
uncertainty.
\end{itemize}

All of the systematic uncertainties are summarised in Table~\ref{BF-Sys}.
Adding them in quadrature results in a total systematic uncertainty of 1.7\%.
\begin{table}[htbp]
\setlength{\abovecaptionskip}{0.cm}
\setlength{\belowcaptionskip}{-0.2cm}
  \begin{center}
    \begin{tabular}{|lc|}
      \hline
      Source   & Uncertainty (\%)\\
      \hline
      Background shape                    & 0.5 \\
      PID efficiency                    & 0.6 \\
      Tracking efficiency               & 1.2 \\
      $\pi^0$ reconstruction              & 0.8 \\
      MC sample size                       & 0.2 \\
      Amplitude model                           & 0.4 \\
      \hline
      Total                               & 1.7 \\
      \hline
    \end{tabular}
  \end{center}
  \caption{Systematic uncertainties in the BF measurement.}
  \label{BF-Sys}
\end{table}

\section{Summary}
%------------------------------------------------------------------------------
The singly Cabibbo-suppressed decay $D_{s}^{+} \to K^+\pi^{+}\pi^{-}\pi^{0}$ is observed, with a BF of $\mathcal{B}(D_s^+ \to K^+\pi^+\pi^-\pi^0) =
(9.75\pm0.54_{\rm{stat.}}\pm0.17_{\rm{syst.}})\times 10^{-3}$.
The first amplitude analysis is also performed, yielding FFs and phases for several significant intermediate states.
Combining these results, we obtain the BFs presented in Table~\ref{inter-processes}.

The dominant intermediate process is determined to be $D_s^+ \to K^{*0}\rho^+$, with a fraction of $(40.5\pm2.8_{\rm{stat.}}\pm1.5_{\rm{syst.}})\%$. The decay $D_{s}^{+} \to K^+\omega$ is observed with a significance greater than 10$\sigma$ and its BF is measured to be $(0.95\pm0.12_{\rm{stat.}}\pm0.06_{\rm{syst.}})\times 10^{-3}$, which is consistent with the BESIII result~$(0.87\pm0.24_{\rm{stat.}}\pm0.08_{\rm{syst.}})\times 10^{-3}$~\cite{LUYU} within 1$\sigma$, but the precision is improved by a factor of 2.1. Information about the two $K_1$ states in this decay provides inputs to further investigations of the mixing between
these two axial-vector kaon states~\cite{PRD93-114010}.
The asymmetry for the BFs of the decays $D_{s}^{+} \to K^+\pi^{+}\pi^{-}\pi^{0}$ and
$D_{s}^{-} \to K^-\pi^{-}\pi^{+}\pi^{0}$ is determined to be $(6.6 \pm 5.4_{\rm{stat.}} \pm 0.7_{\rm{syst.}})$\%.
No evidence for $CP$ violation is found under the current sample size.

\begin{table}[t]
\setlength{\abovecaptionskip}{0.cm}
\setlength{\belowcaptionskip}{-0.2cm}
  \begin{center}
    \begin{tabular}{|lc|}
      \hline
      Intermediate process & BF ($10^{-3}$)\\
      \hline
      $D_s^{+}[S]\to K^*(892)^0\rho^+$  &1.41 $\pm$ 0.23 $\pm$ 0.07 \\
      $D_s^{+}[P]\to K^*(892)^0\rho^+$ &2.53 $\pm$ 0.28 $\pm$ 0.12 \\
      $D_s^{+}\to K^*(892)^0\rho^+$ &3.95 $\pm$ 0.35 $\pm$ 0.17\\
 \hline
      $D_s^{+}[P]\to K^*(892)^+\rho^0$ &0.42 $\pm$ 0.16 $\pm$ 0.06  \\
 \hline
      $D_s^{+}\to K^+\omega$ &0.95 $\pm$ 0.12 $\pm$ 0.06  \\
 \hline
      $D_s^{+}\to K_1(1270)^0\pi^+$, $K_1(1270)^0[S]\to K^+ \rho^-$ &0.39 $\pm$ 0.12 $\pm$ 0.06 \\
 \hline
      $D_s^{+}\to K_1(1400)^0\pi^+$, $K_1(1400)^0[S]\to K^*(892)^+\pi^-$ &0.55 $\pm$ 0.09 $\pm$ 0.03  \\
      $D_s^{+}\to K_1(1400)^0\pi^+$, $K_1(1400)^0[S]\to K^*(892)^0\pi^0$ &0.59 $\pm$ 0.09 $\pm$ 0.02 \\
      $D_s^{+}\to K_1(1400)^0\pi^+$, $K_1(1400)^0[S]\to K^*(892)\pi$ &1.10 $\pm$ 0.19 $\pm$ 0.04 \\
 \hline
      $D_s^{+}\to a_1(1260)^0K^+, a_1(1260)^0[S]\to \rho^+ \pi^-$ &0.19 $\pm$ 0.07 $\pm$ 0.09   \\
      $D_s^{+}\to a_1(1260)^0K^+, a_1(1260)^0[S]\to \rho^- \pi^+$ &0.19 $\pm$ 0.07 $\pm$ 0.09   \\
      $D_s^{+}\to a_1(1260)^0K^+, a_1(1260)^0[S]\to \rho \pi$ &0.32 $\pm$ 0.12 $\pm$ 0.15  \\
 \hline
      $D_s^{+}[S]\to (K^+\pi^0)_{V}\rho^0$ &1.01 $\pm$ 0.20 $\pm$ 0.06  \\
 \hline
      $D_s^{+}\to (K^+\pi^0)_{\rm S-wave}(\pi^+\pi^-)_{\rm S-wave}$ &0.93 $\pm$ 0.22 $\pm$ 0.09  \\
      \hline
    \end{tabular}
  \end{center}
  \caption{The BFs for various intermediate processes
    in the hadronic decay of $D^+_s\to K^+\pi^{+}\pi^{-}\pi^{0}$.
    The first and second uncertainties are
    statistical and systematic, respectively.}
\label{inter-processes}
\end{table}

\acknowledgments
%\vspace{-0.4cm}
The BESIII collaboration thanks the staff of BEPCII and the IHEP computing center for their strong support. This work is supported in part by National Key Research and Development Program of China under Contracts Nos. 2020YFA0406400, 2020YFA0406300; National Natural Science Foundation of China (NSFC) under Contracts Nos. 11625523, 11635010, 11735014, 11775027, 11822506, 11835012, 11875054, 11935015, 11935016, 11935018, 11961141012, 12192260, 12192261, 12192262, 12192263, 12192264, 12192265; the Chinese Academy of Sciences (CAS) Large-Scale Scientific Facility Program; Joint Large-Scale Scientific Facility Funds of the NSFC and CAS under Contracts Nos. U1732263, U1832207, U2032104; CAS Key Research Program of Frontier Sciences under Contracts Nos. QYZDJ-SSW-SLH003, QYZDJ-SSW-SLH040; 100 Talents Program of CAS; INPAC and Shanghai Key Laboratory for Particle Physics and Cosmology; ERC under Contract No. 758462; European Union Horizon 2020 research and innovation programme under Contract No. Marie Sklodowska-Curie grant agreement No 894790; German Research Foundation DFG under Contracts Nos. 443159800, Collaborative Research Center CRC 1044, FOR 2359, FOR 2359, GRK 214; Istituto Nazionale di Fisica Nucleare, Italy; Ministry of Development of Turkey under Contract No. DPT2006K-120470; National Science and Technology fund; Olle Engkvist Foundation under Contract No. 200-0605; STFC (United Kingdom); The Knut and Alice Wallenberg Foundation (Sweden) under Contract No. 2016.0157; The Royal Society, UK under Contracts Nos. DH140054, DH160214; The Swedish Research Council; U. S. Department of Energy under Contracts Nos. DE-FG02-05ER41374, DE-SC-0012069.

\bibliographystyle{JHEP}
\bibliography{references}

\clearpage
\appendix
\section{Clebsch-Gordan relation}
\label{CGrelation}
Considering the isospin relationship in hadron decays, some amplitudes are fixed by Clebsch-Gordan relations, as listed in Table~\ref{tab:CG}. The amplitudes with fixed relations share the same magnitude $(\rho)$ and phase $(\phi)$. 

\begin{table}[t]
  \begin{center}
    \begin{tabular}{|ccc|}
			\hline
      Index                &Amplitude              &Relation\\
      \hline
      $A_1$   &$D_s^{+}\to K_1(1400)^0\pi^+$, $K_1(1400)^0\to K^*(892)^0\pi^0$            & \\
      $A_2$   &$D_s^{+}\to K_1(1400)^0\pi^+$, $K_1(1400)^0\to K^*(892)^+\pi^-$                & \\
      $A$    &$D_s^{+}\to K_1(1400)^0\pi^+$, $K_1(1400)^0\to K^* \pi$                &$A_1-A_2$ \\
			\hline
      $A_1$   &$D_s^{+}\to a_1(1260)^0\pi^+$, $a_1(1260)^0\to \rho^+\pi^-$            & \\
      $A_2$   &$D_s^{+}\to a_1(1260)^0\pi^+$, $a_1(1260)^0\to \rho^-\pi^+$                & \\
      $A$     &$D_s^{+}\to a_1(1260)^0\pi^+$, $a_1(1260)^0\to \rho\pi$                &$A_1-A_2$ \\
     \hline
    \end{tabular}
  \end{center}
  \caption{The Clebsch-Gordan~(CG) relations used.}
    \label{tab:CG}
\end{table}

\renewcommand\thesection{\Alph{section}}
\section{Other intermediate processes tested}
\label{tested_amplitude}
Some other tested amplitudes with significance less than 3$\sigma$ are listed below, the value in each of brackets corresponds to the significance.
\begin{itemize}
\setlength{\itemsep}{0pt}
\item \textbf{Cascade amplitudes}
\item[-] $D_s^{+}\to K^*(892)^+f_0(500),K^*(892)^+\to K^+\pi^0, f_0(500)\to\pi^+\pi^-$~($<1\sigma$)
\item[-] $D_s^{+}\to K^*(892)^+f_0(980),K^*(892)^+\to K^+\pi^0, f_0(980)\to\pi^+\pi^-$~($<1\sigma$)
\item[-] $D_s^{+}[D]\to K^*(892)^0\rho^+,K^*(892)^0\to K^+\pi^-,\rho^+\to\pi^+\pi^0$~($2.6 \sigma$)
\item[-] $D_s^{+}[S,D]\to K^*(892)^+\rho^0,K^*(892)^+\to K^+\pi^0,\rho^0\to\pi^+\pi^-$~($2.0 \sigma$)
\item[-] $D_s^{+}\to \pi^+K_1(1270)^0, K_1(1270)^0[D]\to K^+\rho^-$~($1.5 \sigma$)
\item[-] $D_s^{+}\to \pi^0K_1(1270)^+, K_1(1270)^+[S,D]\to K^+\rho^0$~($<1 \sigma$)
\item[-] $D_s^{+}\to \pi^+K_1(1400)^0, K_1(1400)^0[D]\to K^*\pi$~($2.0 \sigma$)
\item[-] $D_s^{+}\to \pi^+K_1(1650)^0, K_1(1650)^0[S,D]\to K^*\pi$~($2.7 \sigma$)
\item[-] $D_s^{+}\to K^*(1410)^0\pi^+, K^*(1410)^0\to K^*\pi$~($1.9 \sigma$)
\item[-] $D_s^{+}\to K^*(1410)^0\pi^+, K^*(1410)^0\to K^+\rho^-$~($<1 \sigma$)
\item[-] $D_s^{+}\to K^*(1410)^+\pi^0, K^*(1410)^+\to K^*\pi$~($1.9 \sigma$)
\item[-] $D_s^{+}\to K^*(1410)^+\pi^0, K^*(1410)^+\to K^+\rho^0$~($1.8 \sigma$)
\item[-] $D_s^{+}\to K^0(1460)\pi^+, K^0(1460)\to K^*\pi$~($<1 \sigma$)
\item[-] $D_s^{+}\to K^0(1460)\pi^+, K^0(1460)\to K^+\rho^-$~($<1 \sigma$)
\item[-] $D_s^{+}\to K^*(1680)^0\pi^+, K^*(1680)^0\to K^*\pi$~($2.3 \sigma$)
\item[-] $D_s^{+}\to K^*(1680)^+\pi^0, K^*(1680)^0\to K^+\rho^-$~($2.0 \sigma$)
\item[-] $D_s^{+}\to K^+h_1(1170), h_1(1170)[S,D]\to \rho\pi$~($<1 \sigma$)
\item[-] $D_s^{+}\to K^+a_1(1260), a_1(1260)[D]\to \rho\pi$~($<1 \sigma$)
\item[-] $D_s^{+}\to K^+\pi^0(1300), \pi^0(1300)\to \rho\pi$~($2.3 \sigma$)
\item[-] $D_s^{+}\to K^+a_2(1320), a_2(1320)\to \rho\pi$~($1.4 \sigma$)
\item[-] $D_s^{+}\to K^+a_2(1320), a_2(1320)\to \rho(1450)\pi$~($1.5 \sigma$)
\item[-] $D_s^{+}\to K^+\omega(1420), \omega(1420)\to \rho\pi$~($<1 \sigma$)
\item \textbf{Three-body amplitudes}
\item[-] $D_s^{+}[S]\to K^*(892)^+(\pi^+\pi^-)_V$~($2.2 \sigma$)
\item[-] $D_s^{+}[P]\to K^*(892)^+(\pi^+\pi^-)_V$~($2.2 \sigma$)
\item[-] $D_s^{+}[D]\to K^*(892)^+(\pi^+\pi^-)_V$~($2.0 \sigma$)
\item[-] $D_s^{+}[S]\to K^*(892)^0(\pi^+\pi^0)_V$~($2.0 \sigma$)
\item[-] $D_s^{+}[P]\to K^*(892)^0(\pi^+\pi^0)_V$~($<1 \sigma$)
\item[-] $D_s^{+}[D]\to K^*(892)^0(\pi^+\pi^0)_V$~($3.0 \sigma$)
\item[-] $D_s^{+}[S]\to \rho^+(K^+\pi^-)_V$~($<1 \sigma$)
\item[-] $D_s^{+}[P]\to \rho^+(K^+\pi^-)_V$~($1.8 \sigma$)
\item[-] $D_s^{+}[D]\to \rho^+(K^+\pi^-)_V$~($<1 \sigma$)
\item[-] $D_s^{+}[P]\to \rho^0(K^+\pi^0)_V$~($2.0 \sigma$)
\item[-] $D_s^{+}[D]\to \rho^0(K^+\pi^0)_V$~($1.8 \sigma$)
\item[-] $D_s^{+}\to K^*(892)^+(\pi^+\pi^-)_{\rm S-wave}$~($2.0 \sigma$)
\item[-] $D_s^{+}\to K^*(892)^0(\pi^+\pi^0)_{\rm S-wave}$~($1.9 \sigma$)
\item[-] $D_s^{+}\to \rho^+(K^+\pi^-)_{\rm S-wave}$~($2.3 \sigma$)
\item[-] $D_s^{+}\to \rho^0(K^+\pi^0)_{\rm S-wave}$~($<1 \sigma$)
\item \textbf{Four-body non-resonance amplitudes}
\item[-] $D_s^{+}\to K^+((\pi^+\pi^-)_{\rm S-wave}\pi^0)_A$~($<1 \sigma$)
\item[-] $D_s^{+}\to K^+((\pi^+\pi^-)_{\rm S-wave}\pi^0)_P$~($1.6 \sigma$)
\item[-] $D_s^{+}\to \pi^0((\pi^+\pi^-)_{\rm S-wave}K^+)_A$~($1.8 \sigma$)
\item[-] $D_s^{+}\to \pi^0((\pi^+\pi^-)_{\rm S-wave}K^+)_P$~($1.9 \sigma$)
\item[-] $D_s^{+}\to \pi^0((K^+\pi^-)_{\rm S-wave}\pi^+)_A$~($<1 \sigma$)
\item[-] $D_s^{+}\to \pi^0((K^+\pi^-)_{\rm S-wave}\pi^+)_P$~($<1 \sigma$)
\item[-] $D_s^{+}\to \pi^+((K^+\pi^-)_{\rm S-wave}\pi^0)_A$~($2.3 \sigma$)
\item[-] $D_s^{+}\to \pi^+((K^+\pi^-)_{\rm S-wave}\pi^0)_P$~($2.0 \sigma$)
\item[-] $D_s^{+}\to \pi^+((K^+\pi^0)_{\rm S-wave}\pi^-)_A$~($<1 \sigma$)
\item[-] $D_s^{+}\to \pi^+((K^+\pi^0)_{\rm S-wave}\pi^-)_P$~($<1 \sigma$)
\end{itemize}

%\author{Author list}
%\begin{small}
%\begin{center}
\centerline{\large{\textbf{BESIII Collaboration}}}
M.~Ablikim$^{1}$, M.~N.~Achasov$^{11,b}$, P.~Adlarson$^{70}$, M.~Albrecht$^{4}$, R.~Aliberti$^{31}$, A.~Amoroso$^{69A,69C}$, M.~R.~An$^{35}$, Q.~An$^{66,53}$, X.~H.~Bai$^{61}$, Y.~Bai$^{52}$
% ...
\\
%\vspace{0.2cm}
%(BESIII Collaboration)\\
\vspace{0.2cm} {\it
$^{1}$ Institute of High Energy Physics, Beijing 100049, People's Republic of China\\
% ...
$^{4}$ Bochum Ruhr-University, D-44780 Bochum, Germany\\
% ...
$^{11}$ G.I. Budker Institute of Nuclear Physics SB RAS (BINP), Novosibirsk 630090, Russia\\
% ...
$^{31}$ Johannes Gutenberg University of Mainz, Johann-Joachim-Becher-Weg 45, D-55099 Mainz, Germany\\
% ...
$^{35}$ Liaoning Normal University, Dalian 116029, People's Republic of China\\
% ...
$^{52}$ Southeast University, Nanjing 211100, People's Republic of China\\
$^{53}$ State Key Laboratory of Particle Detection and Electronics, Beijing 100049, Hefei 230026, People's Republic of China\\
% ...
$^{61}$ University of Jinan, Jinan 250022, People's Republic of China\\
% ...
$^{66}$ University of Science and Technology of China, Hefei 230026, People's Republic of China\\
% ...
$^{69}$ University of Turin and INFN, (A)University of Turin, I-10125, Turin, Italy; (B)University of Eastern Piedmont, I-15121, Alessandria, Italy; (C)INFN, I-10125, Turin, Italy\\
$^{70}$ Uppsala University, Box 516, SE-75120 Uppsala, Sweden\\
% ...
$^{b}$ Also at the Novosibirsk State University, Novosibirsk, 630090, Russia\\
% ...
}
%\end{center}
%\vspace{0.4cm}
%\end{small}

%\collaborationImg{\includegraphics[width=0.15\textwidth, angle=90]{AT_bes3logo.pdf}}
%\collaboration{BESIII Collaboration}
%\author{\input{author}}
\end{document}